\RequirePackage[T1]{fontenc}
\documentclass[12pt]{article}

\usepackage[height=8.85in,width=6.25in]{geometry}

\usepackage[utf8]{inputenc}
\usepackage{amsmath}
\usepackage{amssymb}
\usepackage{mathtools}
\numberwithin{equation}{section}
\usepackage{slashed}
\usepackage{braket}
\usepackage[svgnames,dvipsnames]{xcolor}
\usepackage[colorlinks,citecolor=DarkGreen,linkcolor=FireBrick,urlcolor=FireBrick,linktocpage,breaklinks=true]{hyperref}
\usepackage{cite}
\usepackage{graphicx}
\usepackage{tikz}
\usetikzlibrary{matrix}

\usepackage{sseq}

\usepackage{times}
\usepackage{courier}
\usepackage{bm}
\usepackage{subfig}
\usepackage{here}
\usepackage{booktabs} 

\usepackage{amsthm}
\theoremstyle{plain}

\DeclareMathOperator{\tr}{tr}

\DeclareMathOperator{\Hom}{Hom}

\DeclareMathOperator{\Sq}{Sq}
\def\CP{\mathbb{CP}}
\def\bR{\mathbb{R}}
\def\bZ{\mathbb{Z}}
\def\bC{\mathbb{C}}
\def\bQ{\mathbb{Q}}
\def\cA{\mathcal{A}}
\def\cP{\mathcal{P}}
\def\ch{\mathrm{ch}}

\begin{document}

\begin{titlepage}

\begin{flushright}

\end{flushright}

\vskip 3cm

\begin{center}

{\Large \bfseries Bosonic SPT and invertible phases \\[1ex]
and its relation to Steenrod's problem}

\vskip 1cm
Shota Saito,
Yuji Tachikawa
and
Yi Zhang
\vskip 1cm

\begin{tabular}{ll}
 & Kavli Institute for the Physics and Mathematics of the Universe (WPI), \\
& University of Tokyo,  Kashiwa, Chiba 277-8583, Japan
\end{tabular}

\vskip 2cm

\end{center}

\noindent 

Bosonic invertible and symmetry-protected topological (SPT) phases are well-known to be 
described by ordinary cohomology groups in low dimensions, 
but `beyond-cohomology' phases appear in higher dimensions.
We make a systematic study of them, and find that the first major non-triviality
is a mod-3 phenomenon, and not a mod-2 phenomenon
as in the case of fermionic  phases.

We also point out that this is a dual manifestation of
the classic question of Steenrod,  namely the issue of the existence of 
homology cycles without manifold representatives.
Thom developed the theory of cobordisms to answer this question,
and we explain how the same analysis leads to Dijkgraaf-Witten phases
which are nontrivial on general simplicial complexes but become trivial on manifolds.

\end{titlepage}

\setcounter{tocdepth}{2}
\tableofcontents

\section{Introduction and summary}
\label{sec:intro}

Symmetry protected topological (SPT) phases are quantum phases which do not possess long-range entanglement, but nevertheless cannot be adiabatically deformed into a trivial product state as long as the relevant symmetry is preserved. The concept of SPT phases was established in \cite{Chen:2011pg,chen2013symmetryprotectedtopologicalorders}, 
where an early proposal was made that $G$-symmetric bosonic SPT phases in $d+1$ dimensions 
would be classified by the elements in $H^{d+1}(G;U_T(1))$ together with their systematic construction.
The same structure was first used by Dijkgraaf and Witten in a different context in the past \cite{Dijkgraaf:1989pz}, and therefore these phases are often called Dijkgraaf-Witten phases. 
The classification of fermionic SPT phases was then pursued in earnest \cite{Gu:2012ib,Kapustin:2014tfa}.
Along the way, it was realized that the classification of SPT phases can be best studied
together with the classification of invertible phases.
Here, invertible phases are a larger class of theories than SPT phases,
and include phases such as the $E_8$ phase, which has nontrivial thermal response
and is not an SPT phase. 
Invertible phases in $d+1$ dimensions are important
in precisely specifying the integrand of the path integral \cite{Freed:2004yc}
and also in characterizing anomalies in $d$ dimensions \cite{Freed:2014iua},
and this connection brought high energy physicists into the study of SPT and invertible phases. Eventually, it was understood that invertible phases can be classified in terms of
Pontryagin or Anderson dual of bordism groups, using stable homotopy theory \cite{Freed:2016rqq,Yonekura:2018ufj}.

In retrospect, the historical development of SPT phases can be organized clearly through the Atiyah-Hirzebruch spectral sequence (AHSS) \cite{Xiong:2016deb,Gaiotto:2017zba}.
The bottom row of the spectral sequence contains the group cohomology data, and 
`beyond-cohomology' phases corresponding to higher rows of the spectral sequence were gradually discovered.
A good example is provided by the early classification of fermionic SPT phases by group supercohomology\cite{Gu:2012ib}.
From the bordism theoretic viewpoint, this can be understood as capturing the bottom two rows of the AHSS for spin bordism.

This point of view also makes it clear that not all bosonic SPT/invertible phases are given by 
ordinary cohomology groups, as was once thought in the early days of the study of these phases.
Indeed, starting in 4+1 dimensions, there are `beyond-cohomology' bosonic phases,
but they are studied only sporadically, e.g.~in \cite{Wan:2018bns,Fidkowski:2019nju,Putrov:2023jqi,Yang:2023gvi,Wan:2026jgg}.
The aim of this paper is to initiate a systematic study of such phases,
and to describe what is to bosonic phases as Gu-Wen supercohomology is to fermionic phases,
putting our focus on the first two nontrivial rows of the corresponding AHSS.
We will find that the first non-triviality is a mod-3 phenomenon
related to the first Pontryagin class $p_1$,\footnote{%
The mod-3 Pontryagin statistics was mentioned also in \cite{Feng:2025mdg} in a different context.
It would be of interest to study possible relationship between that work and our analysis.
}
rather than a mod-2 phenomenon associated to the second Stiefel-Whitney class $w_2$,
as was in the case of fermionic phases.
This issue  manifests first in beyond-cohomology phases in 5+1 dimensions,
as already pointed out by \cite{Yang:2023gvi}. 
We will put it in a more general context.

Closely related to all this is the Steenrod problem, a classical problem in algebraic topology concerning whether a given homology class of a topological space can be realized by a manifold.
It was included in Eilenberg’s list of problems of topology\cite{Eilenberg:1949} as Problem 25, and was solved by Thom in his classic work \cite{Thom1954QuelquesPG}.\footnote{%
It is interesting to note that in \cite{Eilenberg:1949} Steenrod more than one problems,
but somehow now this problem seems to be known as \emph{the} Steenrod problem.
}
As shown there, the first example of a class in $H^n(X;\bZ)$ whose Poincar\'e dual cannot be realized by a manifold occurs when $n=7$.
We will see that this translates to the existence of a Dijkgraaf-Witten phase in $6+1$ dimensions,
which is nontrivial on general simplicial complexes but is trivial on manifolds. 
We write down the action of such a Dijkgraaf-Witten phase with $\bZ_3\times \bZ_3$ symmetry explicitly, 
using the very example Thom considered to settle Steenrod's problem.

The structure of the rest of the paper is as follows.
\begin{itemize}
\item In Section~\ref{sec:2}, we describe various examples of beyond-cohomology bosonic phases,
starting from lower dimensions and gradually going to  higher dimensions.
Along the way, we will give a fairly complete description of these phases in 4+1 dimensions,
and conclude with a concrete description of a Dijkgraaf-Witten phase 
with $\bZ_3\times \bZ_3$ symmetry which is nontrivial on simplicial complexes but is trivial on manifolds.
\item In Section~\ref{sec:3}, we discuss beyond-cohomology bosonic phases systematically.
After giving a brief review of the necessary algebraic topology techniques, 
we explain the examples given in Section~\ref{sec:2}
in terms of the first two nontrivial rows of the AHSS for the oriented bordism groups,
and describe more examples.
In particular, we discuss a phase which is universal in a certain sense in 6+1 dimensions.
\item 
We have three appendices, \ref{app:1}, \ref{app:2} and \ref{app:differentials}.
In Appendix~\ref{app:1} and \ref{app:2}, we give the details of the  computation of the relevant bordism groups,
$\Omega^{SO}_5(B\bZ_3)$ in the former 
and $\Omega^{SO}_8(K(\bZ,4))$ in the latter,
and describe various methods for resolving extension problems in spectral sequences. Finally, in Appendix~\ref{app:differentials}, we review how to determine the
necessary differentials in the AHSS we need to use.
\end{itemize}

Before proceeding, we should mention that we treat invertible/SPT phases from 
the spacetime continuum point of view, only describing their
Euclidean action on closed manifolds.
It would be interesting to study their realizations in Hamiltonian lattice formulations,
as often done in condensed-matter theory.
It would also be interesting to study the boundary theories of the invertible/SPT phases
discussed in this paper, which carry corresponding anomalies.
The authors hope to come back to these questions in the future.

\section{Illustrative examples}
\label{sec:2}
\subsection{The basic 2+1 dimensional phase}\label{sec:The basic 2+1 dimensional phase}
The most fundamental `beyond-cohomology' phase is the 2+1 dimensional phase
described by the gravitational Chern-Simons action \begin{equation}
S= - u  \int_{M_3}  \frac{1}{2\cdot (2\pi)^2} \tr (\Gamma R + \frac23 \Gamma^3).
\label{gravCS}
\end{equation}
Here, $u$ is an overall coefficient, 
$\Gamma$ is the affine connection and $R$ is the metric curvature.
In this paper we use a somewhat unconventional normalization of the action
so that it appears
in the exponentiated Euclidean action as $\exp(2\pi i S)$.
This means that $S$  should be well-defined modulo integers,
i.e.~$S\in \bR/\bZ$. 
We also regard $U(1)$ as an additive group isomorphic to $\bR/\bZ$.

Using the standard linear response theory applied to the action above \eqref{gravCS}, 
it is possible to show that the overall coefficient $u$
is proportional to the thermal Hall conductivity $\kappa_{xy}$,
which measures the energy flow perpendicular to the temperature gradient,
with the precise relation given by  \cite{Read:1999fn,Kapustin:2014dxa} \begin{equation}
\kappa_{xy} = 4\pi u k^2_B T/\hbar.
\end{equation}
The same coefficient $u$ also controls the chiral central charge $c$ of the 1+1 dimensional
boundary theory via the formula \begin{equation}
c = 24 u.
\label{chiralCentral}
\end{equation}

As $\Gamma$ depends on the gauge choice, more care is needed to define it in a topologically nontrivial situation, and this leads to the quantization of the coefficient $u$.
Suppose $M_3$ is closed,
and $W_4$ is a manifold such that $\partial W_4=M_3$. 
Note that \begin{equation}
- d \frac{1}{2\cdot (2\pi)^2} \tr (\Gamma R + \frac23 \Gamma^3 ) 
= -\frac 12 \tr (\frac{R}{2\pi})^2
\end{equation} is the differential form expression of the first Pontryagin class $p_1$ 
of the tangent bundle.
Then we can rewrite \eqref{gravCS}
as \begin{equation}
S=    u  \int_{W_4} p_1.
\end{equation}
This is now independent of the gauge, but possibly depends on the choice of $W_4$.
So let $W'_4$ be another manifold such that $\partial W'_4=M_3$. Then we compare
the two definitions of the exponentiated Euclidean action $e^{2\pi i S}$ using $W_4$ and $W'_4$: \begin{equation}
\frac{\exp( 2\pi i u \int_{W_4} p_1)}{ \exp( 2\pi i u \int_{W'_4} p_1)}
=\exp( 2\pi i u \int_{X_4} p_1),
\end{equation}
where $X_4 = W_4 \sqcup \overline {W'_4}$
is a closed manifold obtained by gluing $W_4$ and the orientation reversal $\overline{W'_4}$
of $W'_4$ along its common boundary $M_3$.

Therefore, we need to demand that $u\int_{X_4} p_1$ is an integer for
arbitrary \emph{oriented} manifolds when our theory is  \emph{bosonic},
whereas we demand the same for arbitrary \emph{spin} manifolds when our theory is \emph{fermionic}.
The Hirzebruch signature theorem says that \begin{equation}
\int_{X_4} \frac{p_1}3
\label{p1/3}
\end{equation}
is equal to the signature of the oriented manifold and hence is an integer,
while the Atiyah-Singer index theorem says that \begin{equation}
\int_{X_4} \frac{p_1}{48}
\end{equation}
is equal to the Dirac index of the spin manifold, and hence is also an integer.
These two quantities can actually take arbitrary integer values,
e.g.~the former is $+1$ on $\CP^2$ and the latter is $-1$ on K3.
Therefore, $u$ is quantized to be an integer multiple of $1/3$  or $1/48$ 
when the theory is bosonic or fermionic, respectively.
Via the relation \eqref{chiralCentral}, this translates to the fact that the chiral central charge of the boundary theory
needs to be a multiple of $8$ or $1/2$ depending on whether the theory is bosonic or fermionic,
and indeed, the basic boundary theory in the bosonic case is the $E_8$ theory with $c=8$,
while the basic boundary theory in the fermionic case is the Majorana-Weyl fermion theory with $c=1/2$.

\subsection{$U(1)$-symmetric 4+1 dimensional phases}\label{sec:U(1)-symmetric 4+1 dimensional phases}
Next let us discuss $U(1)$-symmetric 4+1 dimensional phases.
Let $A$ be the background $U(1)$ gauge field, $F=dA$ be its field strength,
and $c_1=F/(2\pi)$ be the first Chern class. A 5-dimensional oriented manifold $M_5$ with such a $U(1)$-bundle is represented by a pair $(M_5,f )$, where $f: M_5 \rightarrow BU(1)$. The vanishing of $\tilde \Omega^{SO}_5(BU(1))=0$ implies that there exists a six-dimensional $W_6$ with $U(1)$-bundle $\tilde f : W_6 \rightarrow BU(1)$ such that 
\begin{equation}
    \partial W_6 = M_5 \sqcup (\overline M_5)\,, \qquad \tilde f|_{M_5}= f \quad \text{and} \quad \tilde f|_{\overline M_5} = \text{trivial map}\,,
\end{equation}
where $\overline M_5$ is the orientation reversal of $M_5$ and a trivial map from $-M_5$ to $BU(1)$ means a trivial $U(1)$-bundle on $\overline M_5$. Equivalently, $(W_6, \tilde f)$ is the bordism 
\begin{equation}
    (M_5,f)  \sim (M_5, \text{trivial bundle})\,.
\end{equation}
We want to exclude the contribution from the pure gravitational part to the $U(1)$-symmetric phase $S$, hence 
\begin{equation}
    S(M_5, \text{trivial bundle}) = 0\qquad \text{ and }  \qquad S(M_5,f) = \int_{W_6} I(W_6,\tilde f)\,,
\end{equation}
where $I(W_6,\tilde f)$ is some quantity constructed from $W_6$ and the bundle $\tilde f$.  In the following, we use the same symbols $c_1$ and $p_1$ for the characteristic classes on $W_6$ whose restrictions to the boundary give the corresponding classes on $M_5$.

One `cohomological' phase is simply given by \begin{equation}
S=   \int_{W_6} c_1^3.
\label{CS1}
\end{equation} 
That this is independent of the choice of the bordism $(W_6, \tilde f)$ follows from the fact \begin{equation}
\int_{X_6} c_1^3 \in \bZ
\end{equation} for arbitrary closed $X_6$, which in turn follows from the fact that \begin{equation}
c_1^3 \in H^6(BU(1);\bZ)
\label{c1^3}
\end{equation} is naturally a $\bZ$-coefficient cohomology class.

In a classic paper by Dijkgraaf and Witten \cite{Dijkgraaf:1989pz},
a way to write down a $d$-dimensional `Chern-Simons' action for a $G$ gauge field
from an element in $H^{d+1}(BG;\bZ)$ was introduced.
When $G$ is a finite group, $H^{d+1}(BG;\bZ)\cong H^d(BG;U(1))$,
and it is common to refer to these phases for finite $G$ as Dijkgraaf-Witten phases.
We can equally refer to the Chern-Simons terms associated to $H^{d+1}(BG;\bZ)$
for continuous $G$ as Dijkgraaf-Witten phases, but this does not seem to be an
usual practice, and we refrain from doing so.

Now, we saw above that $\int_{X_4}p_1/3$ is always an integer. Then one might be tempted to define a
`beyond-cohomology' phase by \begin{equation}
S\stackrel{?}{=}  \int_{W_6} \frac{p_1}3 c_1.
\end{equation} 
But this does not work, since there are closed manifolds $X_6$ equipped with a $U(1)$ gauge field 
such that \begin{equation}
\int_{X_6} \frac{p_1}3 c_1 \not\in \bZ,
\end{equation}   such as $X_6=\CP^3$ with $c_1$ the standard $U(1)$ bundle 
corresponding to the fibration $S^1\to S^7\to \CP^3$, for which the right hand side is $4/3$.

This issue can be remedied by adding a correction term. Namely, consider the action \begin{equation}
S=   \int_{W_6}(\frac{p_1}3 c_1 -\frac{c_1^3}3).
\label{CS2}
\end{equation}  This action is independent of the choice of $W_6$ bounding $M_5$
because it is known that \begin{equation}
\int_{X_6}(\frac{p_1}3 c_1 -\frac{c_1^3}3) \in \bZ 
\label{U1basic}
\end{equation} for arbitrary closed $X_6$ with a $U(1)$ bundle.
This can be shown either using index theory or algebraic topology.
We will give an algebraic topological argument now;
the index theory will be reviewed later in Sec.~\ref{sec:genind}.

\subsection{An algebraic topology interlude}
In this paper we focus on bosonic theories, and so the fact that the combination \eqref{p1/3} 
or the combination \eqref{CS2} is an integer plays a very important role.
We would therefore like to pause here and provide a derivation,
by reviewing a minimal amount of algebraic topology which will be necessary for us.

\subsubsection{Fermionic phases and the Steenrod squares}
Readers who have studied fermionic invertible phases
might be familiar with Steenrod square operations on $\bZ_2$-coefficient cohomology groups: \begin{equation}
\Sq^i \colon H^k(X;\bZ_2) \to H^{k+i}(X;\bZ_2).
\end{equation} For  $x\in H^k(X;\bZ_2)$, we have 
\begin{equation}
\Sq^i x=0 \quad (k<i), \qquad
\Sq^i x = x^2 \quad (k=i).
\end{equation}
The other two properties we need are
\begin{equation}
\Sq^0 =1, \qquad
\Sq^i (x y) = \sum^i_{k=0} \left(\Sq^k x \right) \left( \Sq^{i-k} y\right) \quad (\text{Cartan formula}).
\end{equation}

Then $\Sq^i x$ for general $k>i$ can be thought of as a generalized version of the square.
Furthermore, the operation \begin{equation}
\beta=\Sq^1 \colon H^k(X;\bZ_2)\to H^{k+1}(X;\bZ_2)
\end{equation} agrees with the Bockstein associated to the sequence 
$0\to \bZ_2\to\bZ_4\to \bZ_2\to 0$.
The Wu formula on closed oriented manifolds says \begin{equation}
\int \beta x=0,  \qquad \int \Sq^2 x = \int w_2 x,
\end{equation} where $w_2$ is the second Stiefel-Whitney class of the tangent bundle.
This relation was the source of the connection between the spin structure, 
governed by the vanishing of $w_2$, and the behavior of the Steenrod squares.

\subsubsection{Oriented phases and the mod-3 Steenrod powers}
For our purposes, Steenrod power operations $\cP^i$ on $\bZ_3$-coefficient cohomology play a similar role in the study of bosonic invertible phases: \begin{equation}
\cP^i \colon H^k(X;\bZ_3) \to H^{k+4i}(X;\bZ_3).
\end{equation} For $x\in H^k(X;\bZ_3)$, we have
\begin{equation}
\cP^i x=0 \quad (k<2i), \qquad
\cP^i x = x^3 \quad (k=2i).
\label{Pi}
\end{equation}
Similar to the Steenrod squares, there are two other properties of $\cP^i$ we need:\begin{equation}
\cP^0 =1, \qquad
\cP^i (x y) = \sum^i_{k=0} \left(\cP^kx \right) \left( \cP^{i-k}y\right) \quad (\text{Cartan formula}).
\end{equation}
In this case, the Bockstein $\beta$ associated to the sequence $0\to \bZ_3\to \bZ_9\to \bZ_3\to 0$
is an operation \begin{equation}
\beta \colon H^k(X;\bZ_3)\to H^{k+1}(X;\bZ_3)
\label{eq:Z3Z9Z3bockstein}
\end{equation} separate from the Steenrod powers $\cP^i$,
and the Wu formula on closed oriented manifolds says \begin{equation}
\int \beta x=0,  \qquad \int \cP^1 x = \int p_1 x, \label{Wu}
\end{equation} where $p_1$ is (the mod-3 reduction of) the first Pontryagin class of the tangent bundle \cite{Hirzebruch}. 

We now have sufficient information to prove that $\int_{M_4} p_1/3$ in Eq.~\eqref{p1/3}, is an integer.
For this, we apply the Wu formula \eqref{Wu} for $x=1\in H^0(M_4;\bZ_3)$,
for which $\cP^1 x =0$ from \eqref{Pi}. Therefore, \begin{equation}
\int_{M_4} p_1 = \int_{M_4} p_1 1 = \int_{M_4} \cP^1 1 = 0 
\end{equation} modulo 3, that is,  $\int p_1$ is a multiple of three.
We can similarly derive that $\int_{M_6}((p_1/3) c_1 - c_1^3/3)$ is an integer.
Indeed, using \eqref{Wu} and \eqref{Pi}, we have \begin{equation}
\int_{M_6} p_1 c_1 = \int_{M_6} \cP^1 c_1 = \int_{M_6}c_1^3
\end{equation} modulo 3, that is, $\int_{M_6} (p_1c_1-c_1^3)$ is a multiple of three.

\subsection{Finite group symmetric 4+1 dimensional phases}\label{sec:Finite group symmetric 4+1 dimensional phases}
\subsubsection{Dijkgraaf-Witten and beyond-cohomology phases}
We now move on  to the 4+1 dimensional phases which are symmetric under a finite group $G$.
The standard `cohomological' phases are the Dijkgraaf-Witten phases \cite{Dijkgraaf:1989pz},
specified by elements \begin{equation}
\alpha \in H^5(BG;U(1)).
\end{equation} These are  analogues of the  $U(1)$-symmetric phase specified by \eqref{c1^3},
since for finite groups we have $H^5(BG;U(1))\cong H^6(BG;\bZ)$.
For a $G$ gauge field specified by the map $f: M_5\to BG$,
the action is simply given by \begin{equation}
\label{eq:DWPinBZ3}
D_{M_5,f}(\alpha):= \int_{M_5} f^*(\alpha) \in \bR/\bZ\cong U(1).
\end{equation}

It is natural to expect that a class of `beyond-cohomology' phases can be specified by 
picking $x\in H^1(BG;U(1))$ and somehow combining it with $p_1/3$,
which is of degree 4 and whose integral is an integer. 
Most naively, we would like to consider \begin{equation}
\int_{M_5} \frac{p_1}{3} f^*(x) ,
\label{verynaive}
\end{equation}
but this is not well-defined,
since a division by three is not well-defined in $U(1)\cong \bR/\bZ$.

We can instead consider  \begin{equation}
\int_{M_5} p_1 f^*(x') \in U(1),\qquad x'\in H^1(BG;U(1))
\label{xnaive}
\end{equation} without trying to divide by $3$.
This does sometimes define a beyond-cohomology phase,
e.g.~when $x'$ comes from $H^1(BG;\bZ_p)$
for a prime $p\neq 3$.
In this case, as $3$ has a multiplicative inverse in $\bZ_p$, 
we can divide \eqref{xnaive} by $3$ consistently,
realizing \eqref{verynaive}.

But this clearly does not work when $p=3$.
Even worse, the phase \eqref{xnaive} is actually zero in that case\footnote{%
This was pointed out recently also in \cite{Wan:2026jgg}.
}, since \begin{equation}
\int_{M_5} p_1 f^*(x')=\int_{M_5} \cP^1 f^*(x')= 0
\label{vanishing}
\end{equation} modulo 3, where we used \eqref{Wu} in the first equality
and \eqref{Pi} in the second equality.
More generally, when $M$ is $n$-dimensional
and $x\in H^{n-4}(BG;\bZ_3)$,
the phase given by $\int_{M_n} p_1 f^*(x)$
is secretly a Dijkgraaf-Witten phase for $y=\cP^1 x$,
since \begin{equation}
\int_{M_n} p_1 f^*(x)=\int_{M_n}  \cP^1 f^*(x)
= \int_{M_n} f^*(\cP^1x )
= \int_{M_n} f^*(y).
\end{equation}

One universal way to combine $x$ with $p_1/3$, which works for any $x$,
is as follows. Note that $f^*(x) \in H^1(M_5;U(1))$
can be thought of as specifying a flat $U(1)$ connection $A$ on $M_5$.
Then we can use the formula \eqref{CS2} to compute its Chern-Simons action:
\begin{equation}
\label{eq:CS5dto6d}
C_{M_5,f}(x) := \int_{W_6} (\frac{p_1}3 y -\frac{y^3}3)
\end{equation} where $y$ is the first Chern class $c_1$ of $A$.
Note that there is no guarantee that the given $G$-bundle on $M_5$ 
can be extended to $W_6$ as a $G$-bundle. Rather, we are extending $f^*(x)$,
which is a flat $U(1)$ gauge field, to $W_6$ as a possibly non-flat $U(1)$ gauge field.

At this point, an attentive reader might point out that we can use \eqref{CS1} to define \begin{equation}
C'_{M_5,f}(x):=\int_{W_6} y^3
\label{y^3}
\end{equation} instead. But this turns out to be an ordinary Dijkgraaf-Witten phase,
due to the following.

Let $A$ be a flat $p$-form gauge field with the holonomy $a\in H^p(M;U(1))$
and the field strength $b=\beta a\in H^{p+1}(M;\bZ)$, where $\beta$  is the Bockstein
associated to the sequence $0\to \bZ\to\bR\to U(1)\to 0$.
Similarly, let $A'$ be a flat $p'$-form gauge field with the holonomy $a'\in H^{p'}(M;U(1))$
and the field strength $b'=\beta a'\in H^{p'+1}(M;\bZ)$.
Assume further that $M$ is of dimension $p+p'+1$.
Then, the Chern-Simons action determined by the class $bb' \in H^{p+p'+2}(W;\bZ)$ 
 can be evaluated without referencing $W$ as \begin{equation}
\int_{W} b b' = \int_{M} a b' = \int_M a\beta a' \in U(1).
\end{equation} 
This can be proved using differential cohomology
(see e.g.~\cite[Sec.~2.2]{Hsieh:2020jpj} for a derivation),
but should be plausible enough for the readers,
because the Bockstein $\beta$ is essentially an exterior derivative $d$,
so $bb'=\beta (a b')$, and $\int_W bb'=\int_M ab'$.

We now apply this to \eqref{y^3} by taking $b'=y^2$, and find \begin{equation}
C'_{M_5,f}(x) = \int_{M_5} f^*(x) y^2 = \int_{M_5}f^*(x(\beta x)^2)
=D_{M_5,f}(x(\beta x)^2).
\end{equation}
Therefore this is simply a Dijkgraaf-Witten phase specified by $x(\beta x)^2 \in H^5(BG;U(1))$.

\subsubsection{A concrete example: $G=\bZ_3$}
\label{sec:Z3}
To familiarize ourselves with this `beyond-cohomology' phase, consider the case $G=\bZ_3$.
Let us take $x\in H^1(B\bZ_3;U(1))=\bZ_3$ to be the generator.
Then $\beta x$ also generates $H^2(B\bZ_3;\bZ)=\bZ_3$,
and similarly $x(\beta x)^2$ generates $H^5(B\bZ_3;U(1))=\bZ_3$.
Then the Dijkgraaf-Witten phase \begin{equation}
D_{M_5,f}(x(\beta x)^2)
\end{equation} is an order-3 invertible phase, which takes the value $1/3$ on $S^5/\bZ_3$.

How about $C_{M_5,f}(x)$? 
Note that \begin{equation}
3C_{M_5,f}(x)=\int_{W_6} (p_1 y -y^3) = \int_{M_5} (p_1 f^*(x) - f^*(x(\beta x)^2)).
\end{equation}
We have already seen that the first term vanishes in \eqref{vanishing}, and therefore \begin{equation}
\label{eq:order9}
3C_{M_5,f}(x)= - D_{M_5,f}(x(\beta x)^2),
\end{equation} meaning that $C_{M_5,f}(x)$ is an order-9 invertible phase.
We note that this phase was previously studied in \cite{Yang:2023gvi}.

\subsubsection{Addition law}

We can also compute $C_{M_5,f}(x+\tilde x)$ in terms of
$C_{M_5,f}(x)$  and $C_{M_5,f}(\tilde x)$ in general. Indeed, we have \begin{align}
C_{M_5,f}(x+\tilde x)
&= \int_{W_6} (\frac{p_1}3(y+\tilde y)-\frac{(y+\tilde y)^3}3) \\
&=C_{M_5,f}(x)+C_{M_5,f}(\tilde x) -\int_{W_6} (y\tilde y^2+\tilde y y^2) \\
&=C_{M_5,f}(x)+C_{M_5,f}(\tilde x) - D_{M_5,f}(x(\beta\tilde x)^2+\tilde x(\beta x)^2).\label{4+1additionlaw}
\end{align}

This means that the totality of Dijkgraaf-Witten phases $D_{M_5,f}(\alpha)$ for $\alpha\in H^5(BG;U(1))$ and
the totality of beyond-cohomology phases $C_{M_5,f}(x)$ for $x\in H^1(BG;U(1))$ 
form an Abelian group $X$ under the stacking of invertible phases, forming an extension \begin{equation}
0\to H^5(BG;U(1))\to X\to H^1(BG;U(1))\to 0, 
\end{equation} whose 2-cocycle $c(x,\tilde x)$ determining the extension 
is given by $x(\beta\tilde x)^2 + \tilde x(\beta x)^2$.
In the explicit case $G=\bZ_3$ we studied above, what we had was \begin{equation}
0\to \bZ_3\to \bZ_9\to \bZ_3 \to 0.
\end{equation}
We will see later that these phases exhaust all the finite $G$-symmetric invertible phases
in 4+1 dimensions, and the Abelian group $X$ is the group they form under stacking.

\subsection{A curious 6+1 dimensional Dijkgraaf-Witten phase}\label{sec:A curious 6+1 dimensional Dijkgraaf-Witten phase}
\label{sec:curious}

The same technique can be used to describe a 6+1 dimensional Dijkgraaf-Witten phase
which is nontrivial on general simplicial complexes but trivial on manifolds.
Let us consider $G=\bZ_3\times \bZ_3$, and use $x$, $\tilde x$ for the two generators of 
$H^1(BG;\bZ_3)=\bZ_3\times \bZ_3$,
and study the Dijkgraaf-Witten phase specified by the element\footnote{%
The computation uses the formula $\beta (xy)=(\beta x)y - x(\beta y)$ for odd degree classes $x$, $y$
and the Cartan formula $\cP^1 (xy)=(\cP^1x) y + x (\cP^1y)$.
}
\begin{equation}
\cP^1 \beta (x\tilde x) = (\beta x)^3 \tilde x - x (\beta \tilde x)^3.
\label{class}
\end{equation}

Strictly speaking, here $\beta = \beta_{\bZ_3\to\bZ_3}$  is the Bockstein in Eq.~\eqref{eq:Z3Z9Z3bockstein} and the Steenrod operation is taken in mod-3 cohomology. Thus $\mathcal P^1\beta(x\tilde x) \in H^7(BG;\mathbb Z_3)$.  The corresponding $U(1)$-valued Dijkgraaf-Witten class is its image 
\begin{equation}
i_{\bZ_3\to\bR/\bZ} \left(\cP^1 \beta (x\tilde x)\right) \in H^7(BG;U(1))
\label{imageclass}
\end{equation} 
under the coefficient homomorphism $i_{\bZ_3\to\bR/\bZ}$ induced by the inclusion $i: \mathbb Z_3\hookrightarrow U(1)$. The expression $(\beta x)^3 \tilde x - x (\beta \tilde x)^3$ is a nonzero $\mathbb Z_3$-class, and its image~\eqref{imageclass} is also nonzero as a $U(1)$-valued class, as we will soon confirm.

Postponing that, let us first show that this Dijkgraaf-Witten phase is always zero on manifolds. 
This is actually true for any class of the form $\cP^1\beta u$.
Indeed, \begin{equation}
\int \cP^1\beta u = \int p_1 \beta u = \int ( \beta (p_1  u) -  (\beta p_1) u )
= -\int (\beta p_1) u
\end{equation} where we used \eqref{Wu} in the first equality,
the Leibnitz rule for the Bockstein in the second equality,
and \eqref{Wu} again in the third equality.
Now, for a mod-3 class $x$, $\beta x$ is zero if  $x$ lifts to an integral class.
$p_1$ is by definition the mod-3 reduction of an integral class, and so $\beta p_1=0$.
Therefore we conclude that \begin{equation}
\int \cP^1 \beta u= -\int (\beta p_1) u = 0. 
\label{eq:vanish}
\end{equation}

Here we note that this computation shows that there can be a homology class
which cannot be realized by manifolds.\footnote{
Here we say a homology class  $[C]$ of a space $X$ is realizable by manifolds
if there is a manifold $M$ and a map $f:M\to X$ such that $[f(M)]=[C]$.
\label{foot:def}
}
To see this, recall that our class \eqref{imageclass} is nonzero.
Dually, there is a 7-cycle $C$ in $BG$, specifying a nonzero element in $[C]\in H_7(BG;\bZ)$,
on which the integral of this class is nonzero.
This argument goes through even if we replace  $B\bZ_3 \times B\bZ_3$,
which is infinite dimensional,
by a finite-dimensional truncation, say $S^7/\bZ_3 \times S^7/\bZ_3$,
since their low-dimensional cohomology groups are the same.
So we have a 7-cycle 
\begin{equation}
[C]\in H_7(S^7/\bZ_3 \times S^7/\bZ_3;\bZ)
\end{equation}
which pairs nontrivially with 
the class 
\begin{equation}
 i_{\bZ_3\to \bR/\bZ}((\beta x)^3 \tilde x - x (\beta \tilde x)^3) \in H^7(S^7/\bZ_3\times S^7/\bZ_3;U(1)).
\end{equation}

This 7-cycle $C$ cannot be realized by manifolds.
Indeed, if it could be realized by a manifold, there would be a manifold $M$
and a map $f:M\to X$.
Then the computation above shows that \begin{equation}
\int_C ( (\beta x)^3 \tilde x - x (\beta \tilde x)^3  )
=\int_C \cP^1 \beta (x\tilde x) 
=\int_M \cP^1 \beta (f^*(x\tilde x)) =0,
\end{equation} leading to a contradiction.
This is actually the very example discussed by Thom in the classic paper \cite{Thom1954QuelquesPG},
see \textit{Un exemple} below Th\'eor\`eme III.9 there.\footnote{%
One can ask a slightly different question, namely, if a homology class $C$ of a manifold $X$
is realized by an embedded submanifold $M\subset X$, not just a manifold $M$ with a map $f:M\to X$.
Clearly, when a class $C$ is realized by an embedded submanifold, 
it is also realized by a manifold in the sense of footnote \ref{foot:def}, since we can take the inclusion $\iota: M \hookrightarrow X$ as the map from $M$ to $X$.
But the converse is not necessarily the case.
It is known, for example, that the generator of $H_7(Sp(2);\bZ)=\bZ$ can be realized by manifolds 
in the sense of footnote \ref{foot:def} but not by any embedded submanifolds \cite{BHK}.
We can also consider immersions $M\looparrowright X$ (which allow self-intersections) instead of just embeddings $M\hookrightarrow X$. 
This question was studied in  \cite{CG}, where it was shown, among others,
that the generator of $H_7(Sp(2);\bZ)=\bZ$ can be realized by an immersion.
This paper \cite{CG} also contains a nice historical overview of the Steenrod problem and other related representability questions.
}

As we recalled in the introduction, Steenrod asked if there is a cycle which cannot be realized by 
manifolds,
and one of the reasons Thom developed the theory of cobordisms was to answer this question.
It is interesting that the same question, phrased dually in the language of cohomology,
predicts the existence of Dijkgraaf-Witten phases which are nontrivial on 
simplicial complexes but trivial on manifolds.
From the point of view of lattice models of topological phases in the condensed matter theory,
it seems perfectly normal to consider simplicial complexes rather than manifolds.
The observation here is that there are topological phases 
that can detect non-manifold-ness of such lattices.
This point might deserve further exploration.

Let us conclude this section by checking that
the element \eqref{imageclass} is actually nonzero. 
 To see this, for a finite $G$, recall the isomorphism $\beta_{\bR/\bZ \to \bZ}:H^7(BG;U(1))\xrightarrow{\sim}H^8(BG;\mathbb Z)$
coming from $0\to \mathbb Z\to \mathbb R\to U(1)\to 0$. 
We only need to show that $\beta_{\bR/\bZ \to \bZ} \circ i_{\bZ_3\to\bR/\bZ}  \left(\cP^1 \beta (x\tilde x) \right)$ is nonzero in $H^8(BG; \bZ)$.
By naturality of Bocksteins,
\begin{equation}
\beta_{\bR/\bZ \to \bZ} \circ i_{\bZ_3\to\bR/\bZ} =\beta_{\bZ_3 \to \bZ}.
\end{equation}
where $\beta_{\bZ_3 \to \bZ}$ is the Bockstein associated to $0\to  \bZ \to  \bZ \to \bZ_3 \to 0$. Let us write
\begin{equation}
Y=\beta_{\bZ_3 \to \bZ} x,\qquad \tilde Y=\beta_{\bZ_3 \to \bZ} \tilde x. 
\end{equation}
Let $\rho_{\bZ \to \bZ_3}$ be the mod 3 reduction, for which there is a compatibility condition 
\begin{equation}
\beta = \rho_{\bZ \to \bZ_3}  \circ \beta_{ \bZ_3 \to \bZ},
\end{equation}
and we have 
\begin{equation}
\rho_{\bZ \to \bZ_3} (Y^3) =  (\beta x)^3 ,\qquad \rho_{\bZ \to \bZ_3} (\tilde Y^3) =  (\beta \tilde x)^3 \,.
\end{equation}
Then we can compute
\begin{equation}
\beta_{ \bZ_3 \to \bZ} \left(\cP^1 \beta (x\tilde x)  \right)
= \beta_{ \bZ_3 \to \bZ} \left( (\beta x)^3 \tilde x - x (\beta \tilde x)^3  \right)=
Y^3\tilde Y-Y\tilde Y^3.
\end{equation}
By the K\"unneth formula for the integral cohomology of $BG=B\mathbb Z_3\times B\mathbb Z_3$, the degree-eight group is
\begin{equation}
H^8(BG;\bZ)
\cong  \bZ_3
\langle
Y^4,\,
Y^3\tilde Y,\,
Y^2\tilde Y^2,\,
Y\tilde Y^3,\,
\tilde Y^4
\rangle
\end{equation}
Thus $Y^3\tilde Y-Y\tilde Y^3\neq 0$. We conclude that the $U(1)$-class~\eqref{imageclass} is nonzero.

\section{Systematic analysis}
\label{sec:3}

We have seen various examples of beyond-cohomology phases.
We have also seen that there are  Dijkgraaf-Witten phases which are nontrivial
on simplicial complexes but trivial on manifolds.
In both cases, the Wu relation $\int \cP^1 x=\int p_1 x$ 
or its generalization $\int \cP^1\beta u =0$ played an important role.
In this section we explain why that was the case,
from the point of view of the Atiyah-Hirzebruch spectral sequence (AHSS)
computing the group of invertible phases.
For mathematical introduction to the necessary materials, textbooks such as \cite{DK,FomenkoFuchs,MillerBook} can be recommended.

\subsection{Generalities}
Beyond-cohomology phases and Dijkgraaf-Witten phases that are nontrivial on simplicial complexes but trivial on manifolds can be understood systematically in terms of the AHSS.
In this subsection, we introduce the AHSS and explain how these phases are understood from this perspective.

The AHSS is a standard tool for computing generalized cohomology and homology theories step by step from ordinary cohomology or homology.
One filters a space by its skeleton, examines the (co)cycles that appear at each stage, and then uses differentials to determine which of them survive, vanish, or become identified.
The remaining data are finally assembled by solving extension problems to obtain the desired generalized cohomology or homology group.

Given a fibration $F\to X\to B$, the AHSSs for $E$-homology and $E$-cohomology 
have the schematic form
\begin{align}
    E^2_{p,q}&=H_p(B;E_q(F))\implies E_{p+q}(X),\quad\ \text{for $E$-homology,} \\
    E_2^{p,q}&=H^p(B;E^q(F))\implies E^{p+q}(X),\quad\text{for $E$-cohomology}.
\end{align}
More specifically, we consider the AHSSs for the bordism theory $\Omega^{SO}_*(X)$ and the invertible phases $\Hom(\Omega^{SO}_*(X),U(1))$ and $(I_\bZ\Omega^{SO})^*(X)$,
since deformation classes of invertible phases in $d+1$ dimensions
are classified by $(I_\bZ\Omega^{SO})^{d+2}(X)$,
whereas invertible phases in $d+1$ dimensions with no 
continuous dependence on the background fields are
classified by $\Hom(\Omega^{SO}_{d+1}(X),U(1))$ \cite{Freed:2016rqq,Yonekura:2018ufj}.

\paragraph{Homology side:}
We use the AHSS for the trivial fibration ${\rm pt}\to X\to X$.
Then, the AHSS for $\Omega^{SO}_d(X)$ takes the form 
\begin{equation} \label{eq:AHSSoriented}
    E^2_{p,q} = H_p(X;\Omega^{SO}_q({\rm pt})) \Longrightarrow  \Omega^{SO}_{p+q}(X) \,.
\end{equation}
We list the groups $\Omega^{SO}_q({\rm pt})$ for $q<10$, which can be found e.g.~in \cite{MilnorStasheff1974}:
\begin{equation} 
    \begin{tabular}{c|cccccccccc} 
  $q$ & $0$ & $1$ & $2$ & $3$ & $4$ & $5$ & $6$ & $7$ & $8$ & $9$ \\ \midrule
$\Omega^{SO}_q({\rm pt})$ & $\bZ$ & $0$ & $0$ & $0$ & $ \bZ $ & $\bZ_2$ & $0$ & $0$ & $\bZ^2$ & $\bZ_2^2$
\end{tabular}\,.
\end{equation}
Therefore the nonzero differential connecting the first two nontrivial
rows have the shape \begin{equation}
d^5_{n,0}\colon E^5_{n,0}\cong H_{n}(X;\bZ)
\to E^5_{n-5,4} \cong H_{n-5}(X;\bZ).
\end{equation}
From general principles, this is given in terms of a stable homology operation.
We specify this using the dual map between Pontryagin dual groups,
\begin{equation}
d_5^{n-5,4}: H^{n-5}(X;U(1)) \to H^n(X;U(1)).
\label{dualdifferential}
\end{equation}
As we will explain in detail below,
this is given  by $i_{\bZ_3\to U(1)} \cP^1 \beta_{U(1)\to \bZ_3}$,
where $i$ changes the coefficient of cohomology classes,
$\cP^1$ is the first mod-3 Steenrod power operation as before,
and $\beta$ is a Bockstein.

Let us study these differentials $d^5_{n,0}$ one by one.
The first one is $d^5_{5,0}:E^5_{5,0}\to E^5_{0,4}$.
But this enters the zeroth column of the spectral sequence.
As for any generalized homology theory,
we have the canonical decomposition 
\begin{equation}
   \Omega^{SO}_i(X) \cong \tilde \Omega^{SO}_i(X) \oplus \Omega^{SO}_i({\rm pt}) \,,
\end{equation}
which means, in terms of AHSS, that the differentials that enter the zeroth columns are necessarily zero.
In particular, $d^5_{5,0}$ is zero.
The next is  $d^5_{6,0}: E^5_{6,0} \to E^5_{1,4}$. However, this map also turns out to be zero, due to a subtler reason.\footnote{
This is because of the following. The Pontryagin dual of the differential 
is $d_5=i\cP^1\beta:H^1(X;\bR/\bZ)\to H^6(X;\bR/\bZ)$.
Take an arbitrary element $u\in H^1(X;\bR/\bZ)$. Then we have $d_5(u)=i_{\bZ_3\to \bR/\bZ}((\beta_{\bR/\bZ\to \bZ_3} u)^3)$.
But $(\beta_{\bR/\bZ\to \bZ_3} u)^3
=\beta_{\bR/\bZ\to \bZ_3} ( u (\beta_{\bR/\bZ\to \bZ} u)^2)$.
Now recall $i_{\bZ_3\to \bR/\bZ}\circ\beta_{\bR/\bZ\to \bZ_3} =0$,
therefore we have $d_5(u)=0$.
This argument crucially relies on the fact that the degree of $u$ is one,
and does not generalize to higher degree elements.}
Therefore, the first possible nontrivial differential can occur only one degree higher, that is $d^5_{7,0}:E^5_{7,0} \to E^5_{2,4}$.

From the above discussion, the $E^2$-page for $\Omega^{SO}_*(X)$ in the first few rows is given as
\begin{equation}
\vcenter{\hbox{
    \begin{sseq}[entrysize=9mm]{9}{6}
        \ssdrop{\bZ}
        \ssmove 0 4 \ssdrop{\bZ}
        \ssmove 0 1 \ssdrop{\bZ_2}
        \ssmoveto 1 0 \ssdrop{*} \ssname{10}
        \ssmove 0 4 \ssdrop{*} \ssname{14}
        \ssmove 0 1 \ssdrop{*}
        \ssmoveto 2 0 \ssdrop{*} \ssname{20}
        \ssmove 0 4 \ssdrop{*} \ssname{24}
        \ssmove 0 1 \ssdrop{*}
        \ssmoveto 3 0 \ssdrop{*} \ssname{30}
        \ssmove 0 4 \ssdrop{*} \ssname{34}
        \ssmove 0 1 \ssdrop{*}
        \ssmoveto 4 0 \ssdrop{*} \ssname{40}
        \ssmove 0 4 \ssdrop{*} \ssname{44}
        \ssmove 0 1 \ssdrop{*}
        \ssmoveto 5 0 \ssdrop{*} \ssname{50}
        \ssmove 0 4 \ssdrop{*} \ssname{54}
        \ssmove 0 1 \ssdrop{*}
        \ssmoveto 6 0 \ssdrop{*} \ssname{60}
        \ssmove 0 4 \ssdrop{*} \ssname{64}
        \ssmove 0 1 \ssdrop{*}
        \ssmoveto 7 0 \ssdrop{*} \ssname{70}
        \ssmove 0 4 \ssdrop{*} \ssname{74}
        \ssmove 0 1 \ssdrop{*}
        \ssmoveto 8 0 \ssdrop{*} \ssname{80}
        \ssmove 0 4 \ssdrop{*} \ssname{84}
        \ssmove 0 1 \ssdrop{*}
        \ssgoto{70}\ssgoto{24}\ssstroke\ssarrowhead
    \end{sseq}
}}
\label{eq:SOAHSS}
\end{equation}
We have drawn the first differential that can potentially be nontrivial:
\begin{equation}
\label{eq:non_zero_diff_in_homology}
    d^5_{7,0}:H_7(X;\bZ)\to H_2(X;\bZ).
\end{equation}

Now we can view Steenrod’s problem in the context of AHSS.
The original problem asked whether, on a given topological space $X$, every homology class $x$ can be represented by a manifold,
in the sense that there is a map $f:M\to X$ from a manifold $M$
such that $x=[f(M)]$.
A homology class is representable in this sense if and only if it lies in the image of $h: \Omega^{SO}_*(X)\to H_*(X;\bZ)$ 
sending $[M,f]$ to $[f(M)]$.
In the language of AHSS, the image of $h$ is $E^\infty_{*,0}$\,: 
\begin{equation}
\Omega^{SO}_*(X) \twoheadrightarrow E^\infty_{*,0} \subset  H_*(X;\bZ),
\end{equation}
which is known as the edge homomorphism. So, for a homology class to be realized by a manifold, it must survive to $E^\infty$. If there is a nontrivial differential $d^5_{*,0}$ and $E^\infty_{*,0}$ is a proper subset of $H_*(X;\bZ)$, this implies the existence of classes that cannot be realized by manifolds.
Again, as mentioned in the introduction,
this was studied by Thom in his foundational paper \cite{Thom1954QuelquesPG}.
\paragraph{Cohomology side:}
Now let us turn to the cohomology side.
We consider two generalized cohomology theories: the Pontryagin dual $\Hom(\Omega^{SO}_*(X),U(1))$, and the Anderson dual $(I_\bZ\Omega^{SO})^*(X)$.
The $E_2$-pages for $\Hom(\Omega^{SO}_*(X),U(1))$ and $(I_\bZ\Omega^{SO})^*(X)$ in the first few rows are given as
\begin{equation}
\vcenter{\hbox{
    \begin{sseq}[entrysize=9mm]{9}{6}
        \ssdrop{\bR/\bZ}
        \ssmove 0 4 \ssdrop{\bR/\bZ}
        \ssmove 0 1 \ssdrop{\bZ_2}
        \ssmoveto 1 0 \ssdrop{*} \ssname{10}
        \ssmove 0 4 \ssdrop{*} \ssname{14}
        \ssmove 0 1 \ssdrop{*}
        \ssmoveto 2 0 \ssdrop{*} \ssname{20}
        \ssmove 0 4 \ssdrop{*} \ssname{24}
        \ssmove 0 1 \ssdrop{*}
        \ssmoveto 3 0 \ssdrop{*} \ssname{30}
        \ssmove 0 4 \ssdrop{*} \ssname{34}
        \ssmove 0 1 \ssdrop{*}
        \ssmoveto 4 0 \ssdrop{*} \ssname{40}
        \ssmove 0 4 \ssdrop{*} \ssname{44}
        \ssmove 0 1 \ssdrop{*}
        \ssmoveto 5 0 \ssdrop{*} \ssname{50}
        \ssmove 0 4 \ssdrop{*} \ssname{54}
        \ssmove 0 1 \ssdrop{*}
        \ssmoveto 6 0 \ssdrop{*} \ssname{60}
        \ssmove 0 4 \ssdrop{*} \ssname{64}
        \ssmove 0 1 \ssdrop{*}
        \ssmoveto 7 0 \ssdrop{*} \ssname{70}
        \ssmove 0 4 \ssdrop{*} \ssname{74}
        \ssmove 0 1 \ssdrop{*}
        \ssmoveto 8 0 \ssdrop{*} \ssname{80}
        \ssmove 0 4 \ssdrop{*} \ssname{84}
        \ssmove 0 1 \ssdrop{*}
        \ssgoto{24}\ssgoto{70}\ssstroke\ssarrowhead
    \end{sseq}
}}
\label{eq:AHSSPontryagin}
\end{equation}
for the Pontryagin dual, and
\begin{equation}
\vcenter{\hbox{
    \begin{sseq}[entrysize=9mm]{9}{7}
        \ssdrop{\bZ}
        \ssmove 0 4 \ssdrop{\bZ}
        \ssmove 0 2 \ssdrop{\bZ_2}
        \ssmoveto 1 0 \ssdrop{*} \ssname{10}
        \ssmove 0 4 \ssdrop{*} \ssname{14}
        \ssmove 0 2 \ssdrop{*}
        \ssmoveto 2 0 \ssdrop{*} \ssname{20}
        \ssmove 0 4 \ssdrop{*} \ssname{24}
        \ssmove 0 2 \ssdrop{*}
        \ssmoveto 3 0 \ssdrop{*} \ssname{30}
        \ssmove 0 4 \ssdrop{*} \ssname{34}
        \ssmove 0 2 \ssdrop{*}
        \ssmoveto 4 0 \ssdrop{*} \ssname{40}
        \ssmove 0 4 \ssdrop{*} \ssname{44}
        \ssmove 0 2 \ssdrop{*}
        \ssmoveto 5 0 \ssdrop{*} \ssname{50}
        \ssmove 0 4 \ssdrop{*} \ssname{54}
        \ssmove 0 2 \ssdrop{*}
        \ssmoveto 6 0 \ssdrop{*} \ssname{60}
        \ssmove 0 4 \ssdrop{*} \ssname{64}
        \ssmove 0 2 \ssdrop{*}
        \ssmoveto 7 0 \ssdrop{*} \ssname{70}
        \ssmove 0 4 \ssdrop{*} \ssname{74}
        \ssmove 0 2 \ssdrop{*}
        \ssmoveto 8 0 \ssdrop{*} \ssname{80}
        \ssmove 0 4 \ssdrop{*} \ssname{84}
        \ssmove 0 2 \ssdrop{*}
        \ssgoto{24}\ssgoto{70}\ssstroke\ssarrowhead
    \end{sseq}
}}
\label{eq:AHSSAnderson}
\end{equation}
for the Anderson dual.

For the Pontryagin dual, the differential is the composition of the Bockstein homomorphism $\beta_{\bR/\bZ\to\bZ_3}$ associated with the following exact sequence
\begin{equation}
    0\xrightarrow[\hspace{3mm}]{}\bZ_3\xrightarrow[\hspace{3mm}]{}\bR/\bZ\xrightarrow[\hspace{3mm}]{\times3}\bR/\bZ\xrightarrow[\hspace{3mm}]{}0,
\end{equation}
the Steenrod operation $\cP^1$, and the embedding $i_{\bZ_3\to\bR/\bZ}$. 

The differential is the one already mentioned in \eqref{dualdifferential}:
\begin{equation}
    d_5=i_{\bZ_3\to\bR/\bZ}\circ\cP^1\circ\beta_{\bR/\bZ\to\bZ_3},
\end{equation}
which we will derive in Appendix~\ref{app:differentials}.

For the Anderson dual, the differential is a map that increases the degree of the cohomology group with $\bZ$ coefficients by five.
It is the composition of the mod-3 reduction $\rho_{\bZ\to\bZ_3}$, the Steenrod operation $\cP^1$, and the Bockstein homomorphism $\beta_{\bZ_3\to\bZ}$ associated with the following exact sequence
\begin{equation}
    0\xrightarrow[\hspace{3mm}]{}\bZ\xrightarrow[\hspace{3mm}]{\times3}\bZ\xrightarrow[\hspace{3mm}]{}\bZ_3\xrightarrow[\hspace{3mm}]{}0.
\end{equation}
Then, the differential is given by 
\begin{equation}
    d_5=\beta_{\bZ_3\to\bZ}\circ\cP^1\circ\rho_{\bZ\to\bZ_3}.
\end{equation}

This will also be determined in Appendix~\ref{app:differentials}.
For simplicity, we will denote the differential for the Pontryagin dual by $\cP^1 \beta$ and the differential for the Anderson dual by $\beta \cP^1$, despite the different meaning of the $\beta$'s. 

On the cohomology side, Steenrod’s problem is formulated as an injectivity problem, while on the homology side it is formulated as a surjectivity problem.
That is, this amounts to asking whether the following maps are injective:
\begin{align}
    H^*(X;U(1))&\to\Hom(\Omega^{SO}_*(X),U(1)),\\
    H^*(X;\bZ)&\to(I_\bZ\Omega^{SO})^*(X).
\end{align}
Physically, this asks whether there exists a nontrivial Dijkgraaf-Witten phase on cycles that becomes trivial when evaluated on manifolds.
In the AHSS, we have the edge homomorphism
\begin{align}
    H^*(X;U(1))& \twoheadrightarrow E_\infty^{*,0}\subset\Hom(\Omega^{SO}_*(X),U(1)),\\
H^*(X;\bZ) &\twoheadrightarrow 
 E_\infty^{*,0}\subset   (I_\bZ\Omega^{SO})^*(X).
\end{align}
Again,  the relevant question for us is whether the differential $d_5$ is nontrivial or not.

\subsection{Examples}

With the method introduced above, in this subsection we would like to rederive the examples treated in section 2 systematically.
We already discussed the AHSS, which is the basic method we use.
But the AHSS determines the groups only up to the extension problem,
which needs to be fixed in some other means.
Here we will use the generalized signature theorem, which we briefly review here.

\subsubsection{Generalized signature theorem}
\label{sec:genind}
It is useful to recall first the relation between the ordinary Dirac index theorem, its twisted version, and the corresponding statement for the signature operator.
Let $D_S$ denote the spin Dirac operator on a spin manifold $M$.
The Atiyah-Singer index theorem gives
\begin{equation}
    \mathrm{ind}D_S=\int_M\hat{A}(TM)\,,
\end{equation}
where $\hat{A}(TM)$ is called the $\hat{A}$-genus  and is a polynomial  of the Pontryagin classes of $M$.

If the Dirac operator is further twisted by a complex vector bundle $E$, then the index is modified by the Chern character $\ch(E)$ of $E$:
\begin{equation}
    \mathrm{ind}D_{S\otimes E}=\int_M\hat{A}(TM)\mathrm{ch}(E)\,,
\end{equation}
where
\begin{equation}
    \ch(E) :=  \tr \left( \exp \left(\frac{i F}{2 \pi}\right) \right) = \text{rank}(E) + \ch_1(E) +  \ch_2(E) + \ch_3(E) + \ldots\,.
\end{equation}
Here the trace is taken in the matrix representation of the curvature $F$ of some connection $A$ on $E$. The $j$-th Chern character $\ch_j(E)$ in the summation has the expression
\begin{equation}
     \ch_j(E) = \frac{1}{j!}\, \tr \left(  \left(\frac{i F}{2 \pi}\right)^j \right)\,.
\end{equation} 
We see that $\ch_j(E)$ is the differential form degree of $2j$.

In the oriented case, the role of the spin Dirac operator is played by the signature operator, 
which is obtained by twisting the spin Dirac operator by the spin bundle.
Let $D_{\mathrm{sign}}$ denote the signature operator on an oriented smooth manifold $M$.
Then the Atiyah-Singer index theorem identifies its index with the Hirzebruch $L$-genus:
\begin{equation}
    \mathrm{ind}D_{\mathrm{sign}}=\int_ML(TM).
\end{equation}
The first few terms of the $L$-genus read
\begin{equation}
    L(TM) = L_0 + L_1 + L_2 + \ldots = 1 + \frac{p_1}{3} + \frac{7 p_2 - (p_1)^2}{45} + \ldots\,,
\end{equation}
where $L_i$ denotes the degree $4i$ part in the expansion.
This index is the ordinary signature of $M$,
that is, $\mathrm{ind}D_{\mathrm{sign}}=\sigma(M)$.

Just as the spin Dirac operator is twisted by a complex vector bundle $E$, the signature operator also admits twisted versions.
For a complex vector bundle $E\to M^n$, the corresponding twisted signature operator has the index
\begin{equation}
\label{eq:gen_ind_theorem}
    \sigma_E(M^n)=\int_{M^n}\sum_{4i+2j=n}L_i2^j\mathrm{ch}_j(E).
\end{equation}
This formula is called the generalized signature theorem. Note the crucial factors $2^j$.
For more details, see \cite[Theorem 3.1.5]{Gilkey1984InvarianceTheory}. 

We now explain how to apply the generalized signature theorem to identify invertible phases.
By definition, the characteristic class appearing on the right-hand side of \eqref{eq:gen_ind_theorem} integrates to an integer,
although they contain non-integer rational coefficients when expanded in terms of
the generators of integral cohomology classes.
This property can be used to solve extension problems in spectral sequences.

\subsubsection{2+1 dimensional phase: $(I_\bZ\Omega^{SO})^4\cong\bZ$}
In the subsection \ref{sec:The basic 2+1 dimensional phase}, we mentioned the gravitational Chern-Simons action in 2+1 dimension.
In this model, the internal symmetry is absent, and as a beyond-cohomology phase it arises from the oriented bordism of a point.
This phase is located in $E_\infty^{0,4}$ in the trivial spectral sequence for the Anderson dual of the oriented bordism as follows:
\begin{equation}
\vcenter{\hbox{
    \begin{sseq}[entrysize=9mm]{1}{5}
        \ssdrop{\bZ}
        \ssmove 0 4 \ssdrop{\bZ}
    \end{sseq}
}}
\label{eq:AHSS_sec2.1}
\end{equation}
To determine the generator of $E^{0,4}_\infty=\bZ$,
note that the signature index theorem \eqref{p1/3} for 4-dimensional closed oriented manifold $M$ says that $\int_M p_1/3$ is an integer.
Therefore $p_1/3$ determines an element of $E^{0,4}_\infty$.
That this is a generator can be checked by noticing that 
the integral of $p_1/3$ on $M=\CP^2$ is 1.
We obtain $E^{0,4}_\infty=\bZ\langle p_1/3\rangle$. 

Physically, this entry at $E^{0,4}$ of the spectral sequence
measures the coefficient of the gravitational Chern-Simons action describing the thermal Hall effect.

\subsubsection{4+1 dimensional phase: $(I_\bZ\Omega^{SO})^6(BU(1))\cong\bZ^2\oplus\bZ_2$}\label{sec:ex2inSA}
In the subsection \ref{sec:U(1)-symmetric 4+1 dimensional phases}, we studied 4+1 dimensional $U(1)$ symmetric phases.
The phases can again be found in the spectral sequence for the Anderson dual of the oriented bordism.
The spectral sequence for the Anderson dual $(I_\bZ\Omega^{SO})^*(BU(1))$ of the oriented bordism is as follows:
\begin{equation}
\vcenter{\hbox{
    \begin{sseq}[entrysize=9mm]{7}{7}
        \ssdrop{\bZ}
        \ssmove 0 4 \ssdrop{\bZ}
        \ssmove 0 2 \ssdrop{\bZ_2}
        \ssmoveto 2 0 \ssdrop{\bZ}
        \ssmove 0 4 \ssdrop{\bZ}
        \ssmove 0 2 \ssdrop{*}
        \ssmoveto 4 0 \ssdrop{\bZ}
        \ssmove 0 4 \ssdrop{*}
        \ssmove 0 2 \ssdrop{*}
        \ssmoveto 6 0 \ssdrop{\bZ}
        \ssmove 0 4 \ssdrop{*}
        \ssmove 0 2 \ssdrop{*}
    \end{sseq}
}}
\label{eq:AHSS_sec2.2}
\end{equation}

We are interested in the part  of $E^{p,q}_\infty$ with $p+q=6$.
The part $E^{0,6}_\infty=\bZ_2$ is canonically split, namely $\left(I_\bZ\Omega^{SO}\right)^6(BU(1)) \cong (I_\bZ\Omega^{SO})^6({\rm pt}) \oplus (\widetilde{I_\bZ\Omega}{}^{SO})^6(BU(1))$, and $E^{0,6}_\infty \cong \left(I_\bZ\Omega^{SO}\right)^6({\rm pt}) \cong \Hom(\Omega_5^{SO}({\rm pt}),U(1)) \cong \bZ_2$ is the purely gravitational part. We do not discuss it any more here.

What matters for us are the two $\bZ$ factors at $E^{2,4}_\infty$ and $E^{6,0}_\infty$.
From this we already know that the reduced group
\begin{equation}
(\widetilde{I_\bZ\Omega}{}^{SO})^6(BU(1))\cong \bZ^2
\end{equation}
as an abstract group, but we would like to determine the two generators.

To do so, note that $E^{2i,0}_2$ is generated by $(c_1)^i$ since the cohomology ring of the classifying space of $U(1)$ is given by $H^*(BU(1);\bZ)\cong\bZ[c_1]$.
Similarly, recall that after tensoring $\bQ$, the generator of $E^{0,4}_\infty$
was $p_1/3$. Therefore, $E^{2,4}_\infty=\bZ\langle c_1p_1/3\rangle$.
Then the  extension problem has the form
\begin{equation}
    0\to \left(E^{6,0}_\infty\right)_{\cong\bZ\langle c_1^3\rangle}\to(I_\bZ\Omega^{SO})^6(BU(1))_{\cong \bZ^2}\to \left(E^{2,4}_\infty\right)_{\cong\bZ\langle c_1p_1/3\rangle}\to0.
\end{equation}

This means that the first generator of $\bZ^2$ is given by the Chern-Simons term
$c_1^3$, which is explained in \eqref{CS1}.
For the second generator, which is `beyond cohomology',
the analysis above means that it has the 
form of $c_1p_1/3+Xc_1^3$ where $X$ is an undetermined constant.
The generalized signature theorem tells us
\begin{equation}
    \sigma_R(M^6)=\int_{M^6}2^3\mathrm{ch}_3(R)+L_12\mathrm{ch}_1(R)=\dfrac{4}{3}c_1(R)^3+\dfrac{2}{3}p_1c_1(R)\in\bZ,
\end{equation}
where $R$ is a representation of $U(1)$, and $\mathrm{ch}(R)$ is the total Chern character of the associated vector bundle for $R$.
Taking $R$ to be the charge-one representation of $U(1)$, we can set $c_1(R)$ equal to the generator of $H^2(BU(1);\bZ)$.
Shifting it further by an integral linear combinations of $c_1p_1$ and $c_1^3$,
we find that the following cohomology class turns out to be an integral class:
\begin{equation}
\label{eq:CS_ex2inSA}
    \dfrac{1}{3}c_1p_1-\dfrac{1}{3}c_1^3.
\end{equation}
This is the other $\bZ$ factor.
Thus, we have seen that the beyond cohomology phase \eqref{CS2} can also be obtained from the spectral sequence.

\subsubsection{4+1 dimensional phase: $\Hom(\tilde{\Omega}^{SO}_5(B\bZ_3),U(1))\cong\bZ_9$}
We constructed in subsection \ref{sec:Z3}
a number of $\bZ_3$-symmetric invertible phases in 4+1 dimensions, by embedding $\bZ_3$ into $U(1)$ and pulling back from $(I_\bZ\Omega^{SO})^6(BU(1))$. 
This corresponds, as described in \eqref{eq:CS5dto6d}, to extending a flat $U(1)$ connection on a five-dimensional manifold to a six-dimensional manifold whose boundary is the given five-dimensional manifold, and then integrating the characteristic class over it.
The pullback along the embedding map $\bZ_3\to U(1)$ is given by
\begin{equation}
    (I_\bZ\Omega^{SO})^6(BU(1))\to(I_\bZ\Omega^{SO})^6(B\bZ_3)\cong\Hom(\Omega^{SO}_5(B\bZ_3),U(1)).
\end{equation}
Therefore, the possible generators on the right-hand side are given by the Chern-Simons terms corresponding to 
\begin{equation}
\label{eq:possibleCS}
    c_1^3,\quad \left(\dfrac{1}{3}c_1p_1-\dfrac{1}{3}c_1^3\right),
\end{equation}
where $c_1$ is the first Chern class of the $U(1)$ bundle, and we ignore the purely gravitational part.
We already saw in subsection \ref{sec:Z3} that they form a $\bZ_9$ group,
but we did not have the ability to show that they exhaust all the possibilities.

Here we do this using the AHSS.

The corresponding spectral sequence for the Pontryagin dual is the following:
\begin{equation}
\vcenter{\hbox{
    \begin{sseq}[entrysize=9mm]{6}{6}
        \ssdrop{\bR/\bZ}
        \ssmove 0 4 \ssdrop{\bR/\bZ}
        \ssmove 0 1 \ssdrop{\bZ_2}
        \ssmoveto 1 0 \ssdrop{\bZ_3}
        \ssmove 0 4 \ssdrop{\bZ_3}
        \ssmoveto 3 0 \ssdrop{\bZ_3}
        \ssmove 0 4 \ssdrop{*}
        \ssmoveto 5 0 \ssdrop{\bZ_3}
        \ssmove 0 4 \ssdrop{*}
    \end{sseq}
}}
\label{eq:AHSS_sec2.4}
\end{equation}
Therefore we have the following exact sequence:
\begin{equation}
\label{eq:ext_AHSS_sec2.4}
    0\to \left(E^{5,0}_\infty\right)_{\cong\bZ_3}\to\Hom(\tilde{\Omega}^{SO}_5(B\bZ_3),U(1))\to \left(E^{1,4}_\infty\right)_{\cong\bZ_3}\to0.
\end{equation}
Therefore the middle term has order 9. We already know that 
it has a $\bZ_9$ subgroup, we conclude that it is actually $\bZ_9$.
The extension was actually already determined in \eqref{eq:order9}.
In addition, we show three other methods to determine the above group extension in Appendix \ref{app:1}.

In conclusion, the Dijkgraaf-Witten phase is an order-3 phase, while the beyond cohomology phase is an order-9 phase.
The entire $\bZ_3$-symmetric invertible phases in 4+1 dimensions is then
generated by  the pullback of the  second Chern-Simons form in \eqref{eq:possibleCS}.
Once again, we see that these phases can be observed in the spectral sequence.
This phase was previously treated in \cite{Yang:2023gvi}, and what we did here was to put it in a more general context.

\subsubsection{5+1 dimensional phase: $(I_\bZ\Omega^{SO})^7(K(\bZ,3))\cong\bZ$}\label{sec:KZ3}
Here we discuss  5+1 dimensional phase with the target space $K(\bZ,3)$.
This target space is associated with a $U(1)$ 1-form symmetry, but we will not go into the physics meaning of this symmetry here.
First, we summarize the cohomology of the Eilenberg-MacLane space $K(\bZ,3)$ in the following:
\begin{equation} 
    \begin{tabular}{c|cccccccccccc}
        $n$ & $0$ & $1$ & $2$ & $3$ & $4$ & $5$ & $6$ & $7$ & $8$ & $9$ & $10$ & $11$ \\
        \midrule
        $H^n(K(\bZ,3);\bZ)$ & $\bZ$ & $0$ & $0$ & $\bZ$ & $0$ & $0$ & $\bZ_2$ & $0$ & $\bZ_3$ & $\bZ_2$ & $\bZ_2$ & $\bZ_3$  \\
        \midrule
        \text{generators} & $1$ &&& $\iota$ &&& $\iota^2$ && $\beta\cP^1 \iota$ & $\iota^3$ & $(Sq^2\iota)^2$ & $\iota\beta\cP^1\iota$
    \end{tabular}
\end{equation}
This data can be obtained e.g.~from $K(\bZ,1)\cong S^1$ by applying the path space fibration twice; it can also be found in \cite{BMT13}.
Note that we only use the data up to degree eight.
In this table, and similarly below, we suppress the coefficient-change maps in order to display the relations among generators.
Steenrod operations are understood to act after reduction modulo the relevant primes, and the displayed torsion generators are identified through the corresponding Bockstein maps.
From this data, we immediately see that there is a nontrivial differential in the AHSS for the Anderson dual of the oriented bordism:
\begin{equation}
\vcenter{\hbox{
    \begin{sseq}[entrysize=9mm]{9}{7}
        \ssdrop{\bZ}
        \ssmove 0 4 \ssdrop{\bZ}
        \ssmove 0 2 \ssdrop{\bZ_2}
        \ssmoveto 3 0 \ssdrop{\bZ}
        \ssmove 0 4 \ssdrop{\bZ} \ssname{34}
        \ssmove 0 2 \ssdrop{*}
        \ssmoveto 5 0 \ssdrop{}
        \ssmove 0 6 \ssdrop{*}
        \ssmoveto 6 0 \ssdrop{\bZ_2}
        \ssmove 0 4 \ssdrop{*}
        \ssmove 0 2 \ssdrop{*}
        \ssmoveto 8 0 \ssdrop{\bZ_3} \ssname{80}
        \ssmove 0 4 \ssdrop{*}
        \ssgoto{34}\ssgoto{80}\ssstroke\ssarrowhead
    \end{sseq}
}}
\label{eq:AHSS_KZ3}
\end{equation}
Recall that the differential in the Anderson dual is given by $\beta\cP^1$, and the nontrivial differential is given by
\begin{equation}
    d_5^{3,4}:E_5^{3,4}\to E_5^{8,0}:\iota\mapsto\beta\cP^1\iota.
\end{equation}
This map is surjective, and therefore, the Anderson dual of the bordism group is given by
\begin{equation}
    (I_\bZ\Omega^{SO})^7(K(\bZ,3))\cong E_\infty^{3,4}\cong3E_2^{3,4} \cong \bZ.
\end{equation}
Recalling that $E_2^{3,4}\cong \bZ$
was generated by $\iota p_1/3$, 
we see that the beyond cohomology phase is given by $\iota p_1$.
It is clear that this integrates to an integer, since both $\iota$ and $p_1$ are integer cohomology classes. 
As a 5+1 dimensional phase, it is simply given by $\int_{M_7} H\wedge p_1$,
where $H = d B$ is the field strength of the $B$-field, which is the background gauge field 
for a $U(1)$ 1-form symmetry. This phase was recently studied in~\cite{Jia:2026jhj}, where the authors apply AHSS directly to $\Omega^{SO}_*(K(\bZ,3))$ and solving extension problem geometrically, obtaining the same results.

\subsubsection{6+1 dimensional phase: $\Hom(\Omega^{SO}_7(B\bZ_3\times B\bZ_3),U(1))\cong\bZ_9^2\oplus\bZ_3^3$}
In the subsection \ref{sec:A curious 6+1 dimensional Dijkgraaf-Witten phase}, we considered a phase that is very different from the previous examples, namely, a phase that can be nontrivial on homology cycles but vanishes when evaluated on manifolds.
This is the Steenrod problem from the cohomological side.
The simplest concrete example is probably the case with $\bZ_3\times\bZ_3$ symmetry in 6+1 dimension, as already given in Thom’s paper\cite{Thom1954QuelquesPG}.

First, we present the relevant spectral sequence of the Pontryagin dual of the oriented bordism group $\Omega^{SO}_*(B\bZ_3\times B\bZ_3)$:
\begin{equation}
\vcenter{\hbox{
    \begin{sseq}[entrysize=9mm]{9}{6}
        \ssdrop{\bR/\bZ}
        \ssmove 0 4 \ssdrop{\bR/\bZ}
        \ssmove 0 1 \ssdrop{\bZ_2}
        \ssmoveto 1 0 \ssdrop{\bZ_3^2}
        \ssmove 0 4 \ssdrop{\bZ_3^2}
        \ssmoveto 2 0 \ssdrop{\bZ_3}
        \ssmove 0 4 \ssdrop{\bZ_3} \ssname{24}
        \ssmoveto 3 0 \ssdrop{\bZ_3^3}
        \ssmove 0 4 \ssdrop{\bZ_3^3} \ssname{34}
        \ssmoveto 4 0 \ssdrop{\bZ_3^2}
        \ssmove 0 4 \ssdrop{*}
        \ssmoveto 5 0 \ssdrop{\bZ_3^4}
        \ssmove 0 4 \ssdrop{*}
        \ssmoveto 6 0 \ssdrop{\bZ_3^3}
        \ssmove 0 4 \ssdrop{*}
        \ssmoveto 7 0 \ssdrop{\bZ_3^5} \ssname{70}
        \ssmove 0 4 \ssdrop{*}
        \ssmoveto 8 0 \ssdrop{\bZ_3^4} \ssname{80}
        \ssmove 0 4 \ssdrop{*}
        \ssgoto{24}\ssgoto{70}\ssstroke\ssarrowhead
    \end{sseq}
}}
\label{eq:AHSS_sec2.5}
\end{equation}
There is a non-trivial differential which is essential in the Steenrod problem:
\begin{equation}
    \cP^1\beta(x\tilde{x})=(\beta x)^3\tilde{x}-x(\beta\tilde{x})^3.
\end{equation}
Therefore, among the five $\bZ_3$ factors of $E_2^{7,0}$, four survive to the $E_\infty$-page.
The $\bZ_3$ factor generated by $i_{\bZ_3\to \bR/\bZ} \left((\beta x)^3\tilde{x}-x(\beta\tilde{x})^3\right)\in H^7(B\bZ_3\times B\bZ_3;U(1))$ is killed by a nontrivial differential.
Therefore, the following edge  homomorphism is not injective:
\begin{equation}
    H^7(B\bZ_3\times B\bZ_3;U(1))\to\Hom(\Omega^{SO}_7(B\bZ_3\times B\bZ_3),U(1)),
\end{equation}
inducing the vanishing of a Dijkgraaf-Witten phase 
\begin{equation}
   i\left( (\beta x)^3\tilde{x}-x(\beta\tilde{x})^3 \right)\mapsto0.
\end{equation}
This means that the above phase vanishes when evaluated on manifolds.
This Dijkgraaf-Witten phase is probably the simplest example of a phase that vanishes on manifolds, in the sense that it involves a small  finite symmetry group and appears in the lowest dimension.
Let us now complete the calculation of the corresponding Pontryagin-dual group.

The remaining possible differential is
\begin{equation}
    d_5^{3,4}:
    \left(E^{3,4}_5\right)_{\cong H^3(B\bZ_3\times B\bZ_3;U(1)) \cong (\bZ_3)^3}
    \rightarrow
    \left(E^{8,0}_5\right)_{\cong H^8(B\bZ_3\times B\bZ_3;U(1))\cong(\bZ_3)^4}.
\end{equation}
A convenient set of generators for $H^3(B\bZ_3\times B\bZ_3;U(1))$ is
\begin{equation}
    i(x\beta x),\qquad i(x\beta\tilde{x})=i(\tilde{x}\beta x),\qquad i(\tilde{x}\beta\tilde{x}).
\end{equation}
Using $d_5=i\cP^1\beta$, we find
\begin{align}
    d_5(i(x\beta x))&=i\cP^1((\beta x)^2)=2i((\beta x)^4)=0,\\
    d_5(i(x\beta\tilde{x}))&=i\cP^1(\beta x\beta\tilde{x})
    =i((\beta x)^3\beta\tilde{x}+\beta x(\beta\tilde{x})^3)=0,\\
    d_5(i(\tilde{x}\beta\tilde{x}))&=i\cP^1((\beta\tilde{x})^2)=2i((\beta\tilde{x})^4)=0.
\end{align}
In the last equalities, the classes inside $i$ are pure polynomials in $\beta x$ and $\beta\tilde{x}$. They are reductions of integral classes, and hence vanish after applying $i:H^8(-;\bZ_3)\to H^8(-;U(1))$.
Therefore $E_\infty^{3,4}\cong E_2^{3,4}\cong(\bZ_3)^3$, and we arrive at the short exact sequence
\begin{equation}
    0\to
    \left(E^{7,0}_\infty\right)_{\cong(\bZ_3)^4}
    \to
    \Hom(\Omega^{SO}_7(B\bZ_3\times B\bZ_3),U(1))
    \to
    \left(E^{3,4}_\infty\right)_{\cong(\bZ_3)^3}
    \to0.
\end{equation}
The extension problem can be solved by using the Adams spectral sequence (ASS).
In Appendix~\ref{appA:adams}, we compute the ASS for a single $B\bZ_3$ and obtain a $\bZ_9$-summand. The two inclusions $B\bZ_3\to B\bZ_3\times B\bZ_3$, together with the corresponding projections, give two such $\bZ_9$-summands. 
Alternatively, the same extension problem can be resolved using the Brown--Peterson method presented in Appendix~\ref{BPmethod}. The two cyclic factors give two copies of the nonsplit extension described in \eqref{eq:BP_Omega7}: two $\bZ_3$-summands in $E^{7,0}_\infty$ combine with two $\bZ_3$-summands in $E^{3,4}_\infty$ to form two $\bZ_9$-summands. We conclude that
\begin{equation}
\begin{split}
        \Hom(\Omega^{SO}_7(B\bZ_3\times B\bZ_3),U(1))\cong \, &\bZ_9\langle \frac{1}{3} i \left(x (\beta x)^3 \right),  \frac{1}{3} i \left(\tilde x (\beta \tilde x)^3 \right)\rangle \\&\oplus\bZ_3 \langle  i \left(x (\beta x)^2 \beta\tilde x \right),   i \left( x \beta  x (\beta \tilde x)^2 \right) , \frac{p_1  }{3} i \left(x \beta\tilde x \right) \rangle,
\end{split}
\end{equation}
where the names we gave for the generators are only meant to be suggestive.
We will see in Sec.~\ref{sec:7dphasesforfinitegroup} that these generators can also be obtained from a universal 6+1 phases for finite groups.

\subsubsection{6+1 dimensional phase: $(\widetilde{I_\bZ\Omega}{}^{SO})^8(K(\bZ,4)),(\widetilde{I_\bZ\Omega}{}^{SO})^8(BG_2)\cong\bZ^2$}\label{sec:exBG2BF4}
Here we consider 6+1 dimensional phases with Lie group symmetry $F_4$ and $G_2$.
Note that the same argument also holds if $F_4$ is replaced by $E_6$, $E_7$ and $E_8$.
The homotopy groups of the two groups are shown below:
\begin{equation} 
    \begin{tabular}{c|cccccccccc} 
        $n$ & $0$ & $1$ & $2$ & $3$ & $4$ & $5$ & $6$ & $7$ & $8$ & $9$ \\
        \midrule
        $\pi_n(BF_4)$ & $0$ & $0$ & $0$ & $0$ & $\bZ$ & $0$ & $0$ & $0$ & $0$ & $\bZ_2$ \\
        \midrule
        $\pi_n(BG_2)$ & $0$ & $0$ & $0$ & $0$ & $\bZ$ & $0$ & $0$ & $\bZ_3$ & $0$ & $\bZ_2$
    \end{tabular}
\end{equation}
By definition, the homotopy group of $K(\bZ,4)$ is $\bZ$ in degree 4 and zero in all other degrees.
Let $f:BG\to K(\bZ,4)$ be the map corresponding to the generator of $H^4(BG;\bZ)\cong\bZ$.
Then $f$ is a $9$-equivalence for $G=F_4$ and a $7$-equivalence for $G=G_2$, see Appendix~A.1 of~\cite{Becker:2025xgy}.
Thus, from now on, we can replace $BF_4$ by $K(\bZ,4)$.
The cohomology of the Eilenberg-MacLane space $K(\bZ,4)$ is given as follows:
\begin{equation} 
    \begin{tabular}{c|cccccccccccc}
        $n$ & $0$ & $1$ & $2$ & $3$ & $4$ & $5$ & $6$ & $7$ & $8$ & $9$ & $10$ 
        \\
        \midrule
        $H^n(K(\bZ,4);\bZ)$ & $\bZ$ & $0$ & $0$ & $0$ & $\bZ$ & $0$ & $0$ & $\bZ_2$ & $\bZ$ & $\bZ_3$ & $0$ 
        \\
        \midrule
        \text{generators} & $1$ &&&& $\iota$ &&& $\beta Sq^2\iota$ & $\iota^2$ & $\beta\cP^1\iota$ 
    \end{tabular}
\end{equation}
Note that, by abuse of notation, we use $\iota$ for the fundamental class, as in the case of $K(\bZ,3)$. 
Since the degree-nine generator is given by the $\beta\cP^1$ operation on the degree-four generator, we see that there is a nontrivial differential in the AHSS of the Anderson dual of the bordism for $K(\bZ,4)$.
The AHSS is given as follows:
\begin{equation}
\vcenter{\hbox{
    \begin{sseq}[entrysize=9mm]{10}{7}
        \ssdrop{\bZ}
        \ssmove 0 4 \ssdrop{\bZ}
        \ssmove 0 2 \ssdrop{\bZ_2}
        \ssmoveto 4 0 \ssdrop{\bZ}
        \ssmove 0 4 \ssdrop{\bZ} \ssname{44}
        \ssmove 0 2 \ssdrop{*}
        \ssmoveto 6 0 \ssdrop{}
        \ssmove 0 6 \ssdrop{*}
        \ssmoveto 7 0 \ssdrop{\bZ_2}
        \ssmove 0 4 \ssdrop{*}
        \ssmove 0 2 \ssdrop{*}
        \ssmoveto 8 0 \ssdrop{\bZ}
        \ssmove 0 4 \ssdrop{*}
        \ssmove 0 2 \ssdrop{*}
        \ssmoveto 9 0 \ssdrop{\bZ_3} \ssname{90}
        \ssmove 0 4 \ssdrop{*}
        \ssgoto{44}\ssgoto{90}\ssstroke\ssarrowhead
    \end{sseq}
}}
\label{eq:AHSS_KZ4}
\end{equation}
The nontrivial differential is surjective and it is given by
\begin{equation}
    d_5^{4,4}:E_5^{4,4}\to E_5^{9,0}:\iota\mapsto\beta\cP^1\iota.
    \label{d544}
\end{equation}
In contrast, the situation is different for $BG_2$.
First, the cohomology of $BG_2$ is as follows:
\begin{equation} 
    \begin{tabular}{c|cccccccccccc}
        $n$ & $0$ & $1$ & $2$ & $3$ & $4$ & $5$ & $6$ & $7$ & $8$ & $9$ & $10$ 
        \\
        \midrule
        $H^n(BG_2;\bZ)$ & $\bZ$ & $0$ & $0$ & $0$ & $\bZ$ & $0$ & $0$ & $\bZ_2$ & $\bZ$ & $0$ & $0$ 
    \end{tabular}
\end{equation}
The difference from $K(\bZ,4)$ is that there is no element in degree nine.
Therefore, in this case, there is no nontrivial differential.
The AHSS for $BG_2$ is given as follows:
\begin{equation}
\vcenter{\hbox{
    \begin{sseq}[entrysize=9mm]{10}{7}
        \ssdrop{\bZ}
        \ssmove 0 4 \ssdrop{\bZ}
        \ssmove 0 2 \ssdrop{\bZ_2}
        \ssmoveto 4 0 \ssdrop{\bZ}
        \ssmove 0 4 \ssdrop{\bZ}
        \ssmove 0 2 \ssdrop{*}
        \ssmoveto 6 0 \ssdrop{}
        \ssmove 0 6 \ssdrop{*}
        \ssmoveto 7 0 \ssdrop{\bZ_2}
        \ssmove 0 4 \ssdrop{*}
        \ssmove 0 2 \ssdrop{*}
        \ssmoveto 8 0 \ssdrop{\bZ}
        \ssmove 0 4 \ssdrop{*}
        \ssmove 0 2 \ssdrop{*}
    \end{sseq}
}}
\label{eq:AHSS_BG2}
\end{equation}

Let us find the generators for both $(\widetilde{I_\bZ\Omega}{}^{SO})^8(K(\bZ,4))\cong\bZ^2$ and $(\widetilde{I_\bZ\Omega}{}^{SO})^8(BG_2)\cong\bZ^2$.
Since there is a nontrivial differential for the case of $K(\bZ,4)$, only a multiple of three of the class remains in the $(4,4)$-component.
Therefore, the generators of the Anderson dual of the bordism consist of the following Dijkgraaf-Witten phase and beyond cohomology phase:
\begin{equation}
    \iota^2,\quad\iota p_1+x\iota^2
\end{equation}
where $x$ is an arbitrary integer constant.
For comparison with the $BG_2$ case, we set $x=-1$.

For $BG_2$, there are only trivial differentials in the range we focus on, so the coefficient $1/3$ of the $(4,4)$-component remains.
Therefore, the phases are as follows:
\begin{equation}
\label{eq:integral_class_G2}
    c_2^2,\quad\dfrac{1}{3}c_2p_1+yc_2^2
\end{equation}
where $c_2$ is the generator of $H^4(BG_2;\bZ)\simeq\bZ$, and $y$ is an unknown rational constant.

To determine the constant $y$, we use the generalized signature theorem in eight dimensions:
\begin{equation}
    \sigma_R(M^8)=\int_{M_8}2^4\mathrm{ch}_4(R)+L_1\,2^2\mathrm{ch}_2(R)\in\bZ
\end{equation}
where $R$ denotes the vector bundle over $M_8$
associated to the $G_2$ representation $R$.

Note that here we have omitted the purely gravitational terms. 
The instanton number $c_2$ is expressed as the characteristic polynomial in the adjoint representation of the group $G$
\begin{equation}
   c_2:=  \frac{1}{4 h_G^\vee} \,\tr_{\mathbf{adj.}}  \!\left(\frac{F}{2 \pi}\right)^2  \,,
\label{eq:instantonnumber}
\end{equation}
and $h_G^\vee$ is the dual Coxeter number for $G$.

Choosing the 7-dimensional vector representation of $G_2$, the degree-4 and degree-8 parts of the  Chern character are given in terms of the instanton number $c_2$ as follows:
\begin{equation}
    \mathrm{ch}_2(\mathbf{7})=-2c_2,\quad\mathrm{ch}_4(\mathbf{7})=\dfrac{1}{6}c_2^2\,,
\end{equation}
where we used $h_{G_2}^\vee=4$ and the trace identities between the vector representation $\mathbf{7}$ and the adjoint representation $\mathbf{14}$ of $G_2$
    \begin{equation*}
    \begin{aligned}
      \tr_{\mathbf{7}}  \! \left(\frac{F}{2 \pi}\right)^2    &= \frac{1}{4} \,\tr_{\mathbf{14}} \! \left(\frac{F}{2 \pi}\right)^2 = 4 \, c_2 \,,\\
         \tr_{\mathbf{7}} \!\left(\frac{F}{2 \pi}\right)^4 &= \frac{1}{4} \left(\tr_{\mathbf{7}}   \!\left(\frac{F}{2 \pi}\right)^2\right)^2 = \frac{1}{4}\left( \frac{1}{4} \tr_{\mathbf{14}}   \!\left(\frac{F}{2 \pi}\right)^2 \right)^2= 4 \left(c_2\right)^2 \,.
    \end{aligned}
    \end{equation*}
Thus, the index is computed by the following formula:
\begin{equation}
    \sigma_{\mathbf{7}}(M^8)=\int_{M_8}\dfrac{8}{3}(c_2^2-c_2p_1),
\end{equation}
which is always an integer.

Since $c_2^2$ and $c_2p_1$ are clearly integral classes, 
we see that \begin{equation}
\frac13(c_2p_1-c_2^2)=\frac{8}{3}(c_2^2-c_2p_1) -3(c_2^2-c_2p_1)
\end{equation} also always integrates to an integer.
This means that we can take  $y=-1/3$ in  \eqref{eq:integral_class_G2}.

Now let us compare the phases obtained in the two cases. We list the generators again.
\begin{equation} 
    \begin{tabular}{c|cc}
        \text{symmetry} & $K(\bZ,4)$ & $BG_2$ \\
        \midrule
        \text{Dijkgraaf-Witten phase} & $\iota^2$ & $c_2^2$ \\
        \text{beyond cohomology phase} & $\iota p_1-\iota^2$ & $(c_2p_1-c_2^2)/3$
    \end{tabular}
\end{equation}
Note that $K(\bZ,4)$ has the same homotopy type as the classifying spaces of the simply connected symmetries $E_6$, $E_7$, $E_8$ and $F_4$ up to sufficiently high degree.
Note that $G_2$ is naturally a subgroup of $F_4$, and $BF_4\simeq K(\bZ,4)$
to the range we are interested. 
Then we can pull back  invertible phases for $K(\bZ,4)$
to get invertible phases for $BG_2$.
The difference in the second row of the table above
  is controlled by the nontrivial differential in \eqref{eq:AHSS_KZ4}.
For $K(\bZ,4)$, this differential removes the would-be $1/3$-normalized generator in the $(4,4)$-component, while for $BG_2$ the corresponding degree-nine target is absent, and therefore this generator is kept.

\subsubsection{6+1 dimensional phase: $(\widetilde{I_\bZ\Omega}{}^{SO})^8(BSpin)$}
As a final example, we consider a 6+1 dimensional system with $Spin(k)$ symmetry, where $k\ge9$.
This phase is interesting because it has a universal property, as we will discuss in the next subsection.
Here, we see some basic facts about how this phase appears in the AHSS.
In this case, the AHSS is given as follows:
\begin{equation}
\vcenter{\hbox{
    \begin{sseq}[entrysize=9mm]{10}{7}
        \ssdrop{\bZ}
        \ssmove 0 4 \ssdrop{\bZ}
        \ssmove 0 2 \ssdrop{\bZ_2}
        \ssmoveto 4 0 \ssdrop{\bZ}
        \ssmove 0 4 \ssdrop{\bZ}
        \ssmove 0 2 \ssdrop{*}
        \ssmoveto 6 0 \ssdrop{}
        \ssmove 0 6 \ssdrop{*}
        \ssmoveto 7 0 \ssdrop{\bZ_2}
        \ssmove 0 4 \ssdrop{*}
        \ssmove 0 2 \ssdrop{*}
        \ssmoveto 8 0 \ssdrop{\bZ^2}
        \ssmove 0 4 \ssdrop{*}
        \ssmove 0 2 \ssdrop{*}
    \end{sseq}
}}
\label{eq:AHSS_BSpin}
\end{equation}
There is no nontrivial differential in this scope.
Since we are not concerned with bordism classes coming from a point, we ignore $(I_\bZ\Omega^{SO})^8(\mathrm{pt})\cong\bZ^2$.
At this point we already know that $(\widetilde{I_\bZ\Omega}{}^{SO})^8(BSpin(k))\cong\bZ^3$, 
and the next step is to identify three generators.
In passing, we note also here that 
the AHSS also tells us that $\Omega^{SO}_7(BSpin(k))\simeq 0$,
which will be needed below.

The $(8,0)$-component of the $E_\infty$ page contains two copies of $\bZ$.
These correspond to the two cohomology phases represented by $(q_1)^2$ and $q_2$.
We denote by $q_1$ the generator of $H^4(BSpin(k);\bZ) \cong \bZ$, normalized as in Eq.~\eqref{eq:instantonnumber}.
Note that the dual Coxeter number for $Spin(k)$ is $h^\vee = k-2$. The standard trace formula for $BSpin(k)$ relates $q_1$ to the quadratic expression in the vector representation $\mathbf{V}$ of $Spin(k)$ as
 \begin{equation}
      \tr_{\mathbf{V}} \!\left(\frac{F}{2 \pi}\right)^2  = \frac{1}{k-2} \,\tr_{\mathbf{adj.}}  \!\left(\frac{F}{2 \pi}\right)^2 = 4 \, q_1 \,.
\end{equation}
We fix the sign convention for $q_2$ by
\begin{equation}
    \tr_{\mathbf V}\!\left(\frac{F}{2\pi}\right)^4=4(q_1)^2-8q_2.
\end{equation}
The fact that $(q_1)^2$ and $q_2$ generate $H^8(BSpin(k);\bZ)\cong\bZ^2$ follows e.g.~from~\cite{BENSON199513}.

The other $\bZ$ in the $(4,4)$-component corresponds to a beyond cohomology phase.
To identify the generator of the phase represented by this class, we again use the generalized signature theorem.
Applying \eqref{eq:gen_ind_theorem} to the vector representation $\mathbf{V}$, we obtain an integral class:
\begin{equation}
    \sigma_{\mathbf{V}}(M^8)=\int_{M^8}\dfrac{8}{3}\{(q_1)^2-2 q_2- p_1q_1\}.
\end{equation}
Here we use
\begin{equation}
    \mathrm{ch}_2(\mathbf{V})=-2q_1,\quad\mathrm{ch}_4(\mathbf{V})=\dfrac{1}{6}(q_1)^2-\dfrac{1}{3}q_2.
\end{equation}
Since $(q_1)^2,q_2,p_1q_1$ are integral and $\sigma_{\mathbf V}(M^8)$ is an integer, the class $\frac{1}{3}\{(q_1)^2-2q_2-p_1q_1\}$ integrates to integers.
Changing the sign and subtracting the integral class $q_2$, we take the generators to be
\begin{equation}
    (q_1)^2,\quad q_2,\quad\dfrac{1}{3}(p_1 q_1- (q_1)^2-q_2).
\end{equation}
The first two are Chern-Simons terms for ordinary cohomology, while the third is the beyond cohomology phase.

\subsection{A `universal' 6+1 dimensional phase}
\subsubsection{Two-stage approximation and $Spin(k)$ symmetry}
So far, we have studied at length the first two layers of the Atiyah-Hirzebruch spectral sequence for the
group of invertible phases, and especially the differential $\beta \cP^1$ connecting the two layers. 
Now, consider the space $P_n$ given by \begin{equation}
K(\bZ,n)\to P_n \to K(\bZ,n -4 )
\end{equation} whose Postnikov invariant $\kappa$ \begin{equation}
P_n \to K(\bZ,n-4)\xrightarrow{\kappa} K(\bZ,n+1)
\end{equation} is given by the cohomology operation \begin{equation}
\beta \cP^1: H^{n-4}(X;\bZ)\to H^{n+1}(X;\bZ). \label{diffP}
\end{equation}
The group $P^n(X):=[X,P_n]$ then has an AHSS
with only two rows  $H^p(X;\bZ)$ at $q=0$ and $q=4$ only,
with the differential given by \eqref{diffP}.
This gives a convenient first `beyond-cohomology' approximation 
to the group $(I_\bZ \Omega^{SO})^n(X)$.
In other words, the two-stage approximation to the group of 
$(d+1)$-dimensional invertible phases parameterized by $X$ is given by $[X, P_{d+2}]$, where $P=\{P_n\}$ is the space defined above.

Here we point out that $P_8$ is 8-equivalent to $BSpin(k)$ for $k \geq 9$.
Let $Q$ be the two-stage Postnikov truncation of $BSpin(k)$ keeping only the first two nontrivial homotopy groups $\pi_4=\bZ$ and $\pi_8=\bZ$.
This truncation then has the form $K(\bZ,8) \to Q \to K(\bZ,4)$.
The possible Postnikov invariants lie in $H^9(K(\bZ,4);\bZ)\cong \bZ_3$, generated by $ \beta \cP^1 \iota $.
In the Serre spectral sequence for this fibration, the generator of $H^8(K(\bZ,8);\bZ)$ transgresses to the Postnikov invariant.
If this invariant were zero, the class $\beta\cP^1\iota$ would survive to $H^9(Q;\bZ)$, giving a $\bZ_3$ summand.
Since $H^9(BSpin(k);\bZ)$ has no such summand, the invariant is nonzero, hence equals $\pm\beta\cP^1\iota$, so $Q\simeq P_8$ for our purposes.

This means that any two-stage oriented invertible phase is a pullback from
a $BSpin(k)$-symmetric oriented invertible phase, which is an eta-invariant for generalized signature operator.
The advantage of this fact is that invertible phases, which had been expressed in terms of abstract Steenrod operations or Postnikov invariants, can be concretely realized as index-theoretic quantities.
This is a fairly concrete (in some definition of the word `concrete') realization of these beyond-cohomology phases, somewhat analogous to the use of $BU(1)$ for 4+1 dimensional phases.

\subsubsection{6+1 dimensional invertible phases for finite groups}
\label{sec:7dphasesforfinitegroup}
Now for a finite group $G$ with a $G$-bundle $f:M_7 \rightarrow BG$, the standard 6+1 dimensional Dijkgraaf-Witten phase is 
\begin{equation}
    D_{M_7,f}(\alpha) := \int_{M_7} f^*(\alpha) \in \bR/\bZ \cong U(1)\,,
    \label{7dD}
\end{equation}
where $\alpha \in H^7(BG;U(1)) \cong H^8(BG;\bZ)$.

We now build beyond-cohomology phases using the $BSpin(k)$-symmetric oriented invertible phase. Following our previous convention, the index-theoretic computation above gives
\begin{equation}
    (q_1)^2, \quad q_2 \quad\text{and}\quad \frac{1}{3}\left(p_1 q_1 - (q_1)^2 - q_2   \right)
\end{equation} as a basis for $\text{Hom} (\tilde \Omega^{SO}_8(BSpin(k)),\bZ)\cong \bZ^3$. 

We now pick a map $g:BG\to BSpin(k)$ (which does not necessarily have to
come from a homomorphism $G\to Spin(k)$).
We then have $g^*(q_1)\in H^4(BG;\bZ)$ and $g^*(q_2)\in H^8(BG;\bZ)$.
Using the isomorphism $\beta: H^n(BG;U(1))\xrightarrow{\simeq} H^{n+1}(BG;\bZ)$
when $G$ is finite, we have elements $x_3 \in H^3(BG;U(1))$ and $x_7 \in H^7(BG;U(1))$ such that \begin{equation}
\beta x_3 = g^*(q_1),\qquad \beta x_7 = g^*(q_2).
\end{equation}
mirroring $q_1$ and $q_2$.
We note that, due to the property of $BSpin(k)$,
we automatically have $\cP^1\beta x_3 =0$.

Now, given a $G$-bundle specified by $f:M_7\to BG$,
we can take a $Spin(k)$ bundle specified by $g\circ f: M_7\to BSpin(k)$.
As $\Omega^{SO}_7(BSpin(k))=0$,
we can take $N_8$ such that $\partial N_8=M_7$
and extend the $Spin(k)$ bundle to $h:N_8\to BSpin(k)$.
We further equip this bundle $h$ with a connection.
We can now write down the Chern-Simons action using these data:
\begin{equation}
    C_{M_7,f}(g) := \int_{N_8}\frac{1}{3} \left( p_1 h^*(q_1) - h^*(q_1)^2 - h^*(q_2) \right)\,.
\end{equation}

Given two maps $g: BG\to BSpin(k)$
and $\tilde g:BG\to BSpin(\tilde k)$,
we have $g\oplus \tilde g: BG\to BSpin(k+\tilde k)$.
The convention for $q_2$ gives
\begin{align}
(g\oplus \tilde g)^* (q_1)&=g^*(q_1)+\tilde g^*(q_1),\\
(g\oplus \tilde g)^* (q_2)&=g^*(q_2)+\tilde g^*(q_2) + g^*(q_1) \tilde g^*(q_1),
\end{align}
and hence
\begin{align}
C_{M_7,f}(g\oplus \tilde g)
&=
C_{M_7,f}(g)
+C_{M_7,f}(\tilde g)
- \int_{N_8} h^*(q_1)\tilde h^*(q_1)\\
&=C_{M_7,f}(g)
+C_{M_7,f}(\tilde g)
-\int_{M_7} f^*(x_3) \beta f^*(\tilde x_3),
\end{align}
where the last term is a Dijkgraaf-Witten phase \eqref{7dD}.

This relation is an analogue of the addition law of 4+1 dimensional bosonic $G$-symmetric
phases given in \eqref{4+1additionlaw},
expressing the sum of two `beyond cohomology' phases 
as a `beyond cohomology' phase plus a Dijkgraaf-Witten phase.
But this unfortunately is not enough to provide the extension cocycle
for $P^8(BG)\cong [BG,BSpin(k)]$ sitting in \begin{equation}
0\to  E^{8,0}_\infty  \to P^8(BG)\cong [BG,BSpin(k)] \to E^{4,4}_\infty\to 0
\end{equation} where \begin{align}
E^{8,0}_\infty &=\mathop{\mathrm{Coker}} \cP^1\beta\colon 
H^2(BG;U(1))\to H^7(BG;U(1)), \\
E^{4,4}_\infty &=\quad\mathop{\mathrm{Ker}} \cP^1\beta\colon
H^3(BG;U(1))\to H^8(BG;U(1)),
\end{align}
since we have not provided an explicit splitting $E^{4,4}_\infty \to [BG,BSpin(k)]$,
i.e.~a way to construct $g:BG\to BSpin(k)$ given a class $x_3\in H^3(BG;U(1))$
satisfying $\cP^1\beta x_3=0$.

\subsection{Toward an analogue of Gu-Wen supercohomology for bosonic phases}
In this paper we have studied 
bosonic invertible and SPT phases in various dimensions
by concentrating on the first two rows. 
An analogous analysis for fermionic phases was pioneered by Gu and Wen \cite{Gu:2012ib}, who introduced their `supercohomology'
to describe the two-stage approximation to the Pontryagin dual
of the spin bordism groups, using our modern understanding.
The way to evaluate it on a spin manifold was developed 
by Gaiotto and Kapustin  \cite{Gaiotto:2015zta},
which was then made mathematically rigorous by Brumfield and Morgan \cite{Brumfiel:2016vpy,BrumfielMorgan4,BrumfielMorganQ}.
It would be nice to have a similar description of the two-stage approximation
to the Pontryagin dual of the oriented bordism.
We do not achieve this in this paper, but let us at least outline 
what needs to be done.

\paragraph{Fermionic case:}
Let us review the case of fermionic phases first. 
The two stage approximation $Q^n(X)$ to $\Hom(\Omega^{Spin}_n(X),U(1))$
is given by \begin{equation}
Q^n(X)= [X,Q_n]
\end{equation} where $Q_n$ sits in the fibration \begin{equation}
    K(\bR/\bZ,n) \rightarrow Q_n \rightarrow K(\bZ_2,n-1)\,,
    \label{eq:Qnfibration}
\end{equation}
with Postnikov invariant $\frac{1}{2} \Sq^2(\iota_{n-1}) \in H^{n+1}(K(\bZ_2,n-1); \bR/\bZ)$, where the $\frac{1}{2}$ is the natural map $H^\bullet(-;\bZ_2) \to H^\bullet(-;\bR/\bZ)$.

A class in $Q^n(X)$ then has the form $[a,b]$ for \begin{equation}
 (a,b) \in C^{n}(X;\bR/\bZ) \times C^{n-1}(X;\bZ_2)\,,
\end{equation} satisfying \begin{equation}
    \delta a = \frac{1}{2} \Sq^2 b\,, \quad \delta b= 0\,
\end{equation} with a suitable equivalence relation imposed.
Then the group law is given by \begin{equation} 
    [(a,b)] + [(\tilde a, \tilde b)] = \left[ \left(a+ \tilde a + \frac{1}{2} b \cup_{n-2} \tilde b, b + \tilde b \right) \right]\,,
    \label{eq:twolayerspingrouplaw}
\end{equation} where the extension cocycle term corrects the lack
of the additivity of $\Sq^2$ at the level of cochains:
\begin{equation}
\delta ( \frac{1}{2} b \cup_{n-2} \tilde b)=
\frac{1}{2} \Sq^2  (b+\tilde b)
-\frac{1}{2} \Sq^2  b - \frac{1}{2} \Sq^2 \tilde b.
\end{equation}

One then wants to evaluate the pair $[a,b]$ on a closed spin $n$-manifold $M$ with a map to $X$.
The naive expression 
\begin{equation}
    \int_{M} f^*(a) \,,
\end{equation}
can not work, since $a$ is not necessarily closed to start with. 

One can try to add a correction term $-q_M\left(f^*(b)\right)$ that also depends on $b$.   
Then the combination  
\begin{equation}
    E_M(a,b):=\left(\int_{M} f^*(a)\right)- q_M\left(f^*(b)\right)  \in  \bR/\bZ \cong U(1)
\end{equation}
should be invariant under the change of representatives of $[(a,b)]$ and should be compatible with the group law~\eqref{eq:twolayerspingrouplaw}. 
The construction of such a $q_M$ was indicated in \cite{Gu:2012ib},
made more precise in \cite{Gaiotto:2015zta},
and then was further studied in \cite{Brumfiel:2016vpy,BrumfielMorgan4,BrumfielMorganQ}.

\paragraph{Bosonic case:}
For the bosonic case, 
the two stage approximation $R^n(X)$ to $\Hom(\Omega^{SO}_n(X),U(1))$
is given by \begin{equation}
R^n(X)= [X,R_n]
\end{equation} where $R_n$ sits in the fibration \begin{equation}
    K(\bR/\bZ,n) \rightarrow R_n \rightarrow K(\bR/\bZ,n-4)\,,
    \label{eq:Rnfibration}
\end{equation}
whose Postnikov invariant is our favorite\begin{equation}
\cP^1 \beta : H^{n-4}(X;\bR/\bZ)\to H^{n+1}(X;\bR/\bZ).
\end{equation}

A class in $R^n(X)$ would then has the form $[a,b]$ for \begin{equation}
 (a,b) \in C^{n}(X;\bR/\bZ) \times C^{n-4}(X;\bR/\bZ)\,,
\end{equation} satisfying \begin{equation}
    \delta a = \cP^1\beta b\,, \quad \delta b= 0\,
\end{equation} with a suitable equivalence relation imposed.
Then the group law is given by \begin{equation} 
    [(a,b)] + [(\tilde a, \tilde b)] = \left[ \left(a+ \tilde a + c(b,\tilde b), b + \tilde b \right) \right]\,,
    \label{eq:twolayerbosonicgrouplaw}
\end{equation} where the extension cocycle term should correct the lack
of the additivity of $\cP^1\beta $ at the level of cochains:
\begin{equation}
\delta c(b,\tilde b)=
\cP^1\beta  (b+\tilde b)
-\cP^1 \beta b - \cP^1\beta \tilde b.
\end{equation}
To make these explicit, we need to have an explicit cochain-level expression
for the Steenrod power $\cP^1$,
which was recently achieved in \cite{MM1,MM2},
which should be helpful.

One then wants to evaluate the pair $[(a,b)]$ on a closed oriented $n$-manifold $M$ with a map to $X$.
Following the analogies with the fermionic case, 
we would need an expression of the form
\begin{equation}
    E_M(a,b):= \left(\int_{M} f^*(a)\right)- \xi_M\left(f^*(b)\right)  \in  \bR/\bZ \cong U(1),
\end{equation}
and the construction of  a suitable $\xi_M\left(f^*(b)\right)$ lies at the center of the question.
We would like to know how this $\xi_M$ is related to the naive
formula \begin{equation}
\int_M \frac{p_1}3 f^*(b) \label{bey}
\end{equation}
 for the `beyond cohomology' phase.

We do not have a formula for $\xi_M$ yet,
but we can say the following.
To lighten the notation,
we drop the pullback symbol $f^*$ from now on.
We have a map \begin{equation}
C^{n-4}(X;\bR/\bZ) \xrightarrow{3\times} C^{n-4}(X;\bR/\bZ).
\end{equation}
We pick a section of this map and denote it by $b\mapsto \frac b3$,
by a slight abuse of notation,
and use $\int_M p_1\frac b3$ as a substitute for \eqref{bey}.
We are thus led to consider the combination \begin{equation}
\int_M (a- p_1 \frac b3).
\end{equation} Suppose $M$ and $M'$ are related by a bordism $N$.
Then \begin{equation}
\int_M (a- p_1 \frac b3)
-\int_{M'} (a- p_1 \frac b3)
=\int_N (\delta a-p_1\beta b)
=\int_N( \cP^1 \beta b -p_1\beta b).
\end{equation}
As $\cP^1\beta b-p_1\beta b$ integrates to zero on closed manifolds
thanks to the Wu formula,
hopefully there would be a correction cochain $\zeta(b)$,
universally constructed from various data, such that \begin{equation}
\delta\zeta (b) = \cP^1 \beta b -p_1\beta b.
\end{equation}
If so, \begin{equation}
\xi_M(b):=\int_M (p_1 \frac{b}3  -\zeta(b))
\end{equation} would do the job.
If this line of thought is on the right track, what would be necessary is to
construct  $\zeta(b)$,
which can be thought of as a cochain-level witness of our crucial Wu formula.
The authors would hope to come back to this question in the future.

\section{Summary and outlook}

Let us close by spelling out the picture that emerged from the computations above.
In the first two rows of the AHSS, the oriented theory runs parallel to the spin theory, but with the mod-$3$ operation $\beta\cP^1$ playing the role played by $\Sq^2$ in the spin case.
This difference is tied to the first Pontryagin class $p_1$, and
the fact that $p_1/3$ integrates to integers on oriented manifolds.
This factor of 3 was responsible for the first bosonic beyond-cohomology phenomena discussed in this paper.

The same perspective also makes the relation of our findings with the Steenrod problem explicit.
The AHSS differential produces Dijkgraaf-Witten phases which are nontrivial on general spaces but become trivial when evaluated on manifolds.
The 6+1 dimensional phase with $\bZ_3\times\bZ_3$ symmetry gives an explicit example of this phenomenon, and can be viewed as a physical counterpart of the failure of certain homology classes to be represented by manifolds.

For 4+1 dimensional phases, we saw that two $U(1)$-symmetric invertible phases,
given by Chern-Simons terms for $c_1^3$ and $(p_1 c_1-c_1^3)/3$,
can be used to produce all finite-group symmetric phases,
and to determine the addition law satisfied by them.
By pulling these phases back along $\bZ_3 \to U(1)$, we found that
the beyond-cohomology phase with $\bZ_3$ symmetry has order 9, for example.

For 6+1 dimensional phases, we studied $G_2$-symmetric and $F_4$-symmetric phases. They are closely related, but showed different behavior in the AHSS,
because the differential in the spectral sequence acting on the 
generator of $H^4(BG_2;\bZ)=\bZ$ was trivial, while the differential acting on the generator of $H^4(BF_4;\bZ)=\bZ$ was nontrivial.
This gave a factor of $3$ difference in the normalization of allowed beyond-cohomology phases.
We also saw how this can be reproduced from the perspective of the generalized signature theorem.

For 6+1 dimensional phases, we also found that the comparison with $BSpin(k)$ for $k>9$ gave a useful general way to realize the same two-stage approximation.
This was because the two-stage Postnikov truncation of $BSpin(k)$ is equivalent to the two-stage truncation for the Anderson dual of the oriented bordism group.

In this paper we concentrated on the first two nontrivial stages of the AHSS for the bosonic invertible phases.
For the spin case, the Gu-Wen supercohomology gives a perfectly general cochain-level understanding of the first two stages.
Furthermore, it is known how to evaluate such a cochain-level representative on a spin manifold equipped with a background field for the symmetry group.
The corresponding general construction for the bosonic invertible phases is
still missing.
For this purpose, we would need a concrete cochain representative for the crucial differential $\cP^1\beta$, 
together with a correction term whose coboundary realizes the relevant Wu formula at the cochain level.

Another glaring omission in this paper is that we have only considered 
the Euclidean action on general manifolds, and have not discussed at all
the Hamiltonian lattice perspective. 
As one of the core issues we identified in this paper was 
the existence of the Dijkgraaf-Witten phase which is nontrivial on general 
simplicial complexes but is trivial on manifolds,
it would be very interesting to understand how this could be understood in the Hamiltonian lattice construction.

Finally, it might be of interest to try to incorporate more rows of the AHSS into the analysis.
It is known \cite{Troue}  that the Postnikov invariant of the oriented bordism spectrum is never 2-torsion.
This means, for example, that the next nontrivial row, giving $\Omega^{SO}_5({\rm pt})=\bZ_2$,
simply contributes a decoupled copy of $H_p(X;\bZ_2)$ to $\Omega^{SO}_{5+p}(X)$.
This fact might already be of some interest; but this also means that, to see some further nontrivial behaviors,
we need to go to even higher rows.

The authors hope to return to these and other questions in the future.

\section*{Acknowledgments}
YT thanks illuminating discussions with Ryohei Kobayashi, Kazuaki Takasan, and Mayuko Yamashita.
The authors thank Philip Boyle Smith for explaining how to solve extension problems using the Adams spectral sequence.

SS is supported by Forefront Physics and Mathematics Program to Drive Transformation (FoPM), a World-leading Innovative Graduate Study (WINGS) Program, the University of Tokyo, by Research Fellow of Japan Society for the Promotion of Science (JSPS Research Fellow), by JSPS KAKENHI Grant Number JP25KJ0857, and by JSR Fellowship, the University of Tokyo.
YT is supported in part by WPI Initiative, MEXT, Japan at Kavli IPMU, the University of Tokyo
and by JSPS KAKENHI Grant-in-Aid (Kiban-C), No.24K06883. YZ is supported by WPI Initiative, MEXT, Japan at Kavli IPMU, the University of Tokyo.

\appendix

\section{The computation of $\tilde{\Omega}^{SO}_5(B\bZ_3)$}
\label{app:1}

In this section, we study $\tilde{\Omega}^{SO}_5(B\bZ_3)$ using three complementary methods.\footnote{%
We note that $\Omega^{SO}(B\bZ_n)$ was determined in one of the  very books which started the study of oriented bordism groups \cite{ConnerFloydBook}, whose second edition is \cite{ConnerBook}.
See e.g.~Chapter III in the second edition.
Note that the terminologies and the techniques in this book were rather old fashioned, but that the book still contains useful and instructive materials.
}
The point is to resolve the extension problem left open by the AHSS calculation in the main text. 
Namely, the AHSS for its Pontryagin dual is presented in \eqref{eq:AHSS_sec2.4} in the main text.
The extension problem is given by \eqref{eq:ext_AHSS_sec2.4}, and the bordism group is left with the two possibilities $\bZ_3\times\bZ_3$ and $\bZ_9$.

The three methods we describe are:
(i) computing the eta invariant on spherical space forms, which we present in Appendix~\ref{appA:eta},
(ii) computing the Adams spectral sequence (ASS) in the exterior Hopf subalgebra generated by the Milnor primitives, which we present in Appendix~\ref{appA:adams}, and
(iii) using the $BP$-decomposition of the $MSO$ spectrum, which we explain in Appendix~\ref{BPmethod}.
These computations as examples may be useful when the extension problems in the AHSS or the Adams spectral sequence (ASS) in other sub algebras, which is often allows us to determine extension problems in AHSS for spin bordism, for some other bordism groups cannot be fully resolved.

These three subsections can be read completely separately. 
Throughout this appendix, we focus on the reduced group, namely the symmetry-dependent part. The unreduced group also contains the point contribution $\Omega^{SO}_5(\mathrm{pt})\cong\bZ_2$, generated by the Wu-manifold $SU(3)/SO(3)$.

\subsection{Using eta invariants of spherical space forms}
\label{appA:eta}
Here, we consider a method for solving the extension problem using spherical space forms.
The strategy is to consider a lens space and compute the eta invariant on it. As the eta invariant of the signature operator twisted by
$\bZ_3$-bundles on five-dimensional manifolds is a bordism invariant,
if a phase of $1/9$ appears, the extension is non-split and we have $\bZ_9$.
This method is not foolproof; it might just produce a phase of $1/3$ rather than $1/9$.
Then one needs to try some other manifolds with $\bZ_3$-bundles.
Still, there is some advantage in this method, since  there is an explicit formula for the eta invariant, it can be computed quite easily,
and there can be many spherical space forms.

A spherical space form is a manifold whose universal cover is a sphere.
More concretely, it is a quotient of a sphere by a free action of a finite group.
One often takes this action to be by isometries, so that a spherical space form can be written as
\begin{equation}
    M\cong S^n/\Gamma
\end{equation}
where $\Gamma$ is a finite subgroup of $O(n+1)$ acting freely on $S^n$.

Lens spaces are the simplest and most familiar examples.
A class in $\Omega^{SO}_5(B\bZ_3)$ is represented by a map $M^5\to B\bZ_3$, where $M^5$ is a closed five-dimensional manifold.
The classifying space is $B\bZ_3\cong K(\bZ_3,1)\cong S^\infty/\bZ_3$.
In the standard infinite lens space model $S^\infty/\bZ_3$, the five-skeleton is $S^5/\bZ_3$.
Thus, the inclusion $S^5/\bZ_3\hookrightarrow B\bZ_3$ defines a class in $\Omega^{SO}_5(B\bZ_3)$, and we use the same notation for its reduced class in $\tilde{\Omega}^{SO}_5(B\bZ_3)$.\footnote{Recall that $\Omega^{SO}_n(X)$ splits as $\tilde{\Omega}^{SO}_n(X)\oplus\Omega^{SO}_n(\mathrm{pt})$. The projection to the reduced group sends $[M,f]$ to $[M,f]-[M,\mathrm{triv}]$, where $\mathrm{triv}$ is the constant map to $X$. This is the same consideration as in Section~\ref{sec:U(1)-symmetric 4+1 dimensional phases}. We use this projection when writing a map $M\to X$ as a reduced class.}
Fortunately, the formula for the eta invariant that we need is explicit and is given in part (a) of Lemma 2.1 of \cite{Gilkey1984}.
The eta invariant is given as
\begin{equation}
    \eta([S^5/\bZ_3,S^5/\bZ_3\hookrightarrow B\bZ_3])=\dfrac{1}{3}\left\{\omega\left(\dfrac{\omega+1}{\omega-1}\right)^3+\omega^2\left(\dfrac{\omega^2+1}{\omega^2-1}\right)^3\right\}=-\dfrac{1}{9}
\end{equation}
where $\omega$ is a primitive third root of unity, i.e. $\omega^3=1$ and $\omega \neq 1$.
Since this pairing has denominator $9$, the reduced class on which the eta invariant is evaluated has order $9$. Together with the AHSS calculation, this implies $\tilde{\Omega}^{SO}_5(B\bZ_3)\cong\bZ_9$, and $[S^5/\bZ_3,S^5/\bZ_3\hookrightarrow B\bZ_3]$ is the geometric generator of the reduced group.

\subsection{Using Adams spectral sequence}
\label{appA:adams}
Here, we demonstrate how the ASS resolves the extension problem.\footnote{
The authors thank Philip Boyle Smith for pointing out that the extension problem can be resolved in ASS.
}
In many cases, it is recommended to use both the ASS and the AHSS when computing bordism groups.
For a detailed account of the computation of the ASS, we refer the reader, for example, to \cite{Beaudry:2018ifm,Wan:2018bns}.
The ASS in this case is given by
\begin{equation}
\label{eq:ASS_BZ3}
    \mathrm{Ext}_{\mathcal{A}_3}^{s,t}(\widetilde{H}^*(MSO\wedge B\bZ_3;\bZ_3),\bZ_3)\implies\widetilde{\Omega}_{t-s}^{SO}(B\bZ_3)_3^\wedge.
\end{equation}
where $\mathcal{A}_3$ is the mod $3$ Steenrod algebra, $MSO$ is the  Thom spectrum of the universal bundle over $BSO$, and the subscript/superscript ${}_3^\wedge$ means the 3-completion.
As an $\mathcal{A}_3$-module, the $\bZ_3$ cohomology of $MSO$ is given by
\begin{equation}
    H^*(MSO;\bZ_3)\cong\mathcal{A}_3/(\beta)\oplus M_{\ge8},
\end{equation}
where $(\beta)$ is the two-sided ideal generated by the Bockstein homomorphism, and $M_{\ge8}$ is an undetermined module of degree eight or higher.
Since we are interested only in degrees up to seven, we consider only $\mathcal{A}_3/(\beta)$ from this point onward.
This can also be seen from the fact that the 3-localized $MSO$ spectrum splits as a wedge sum of suspensions of the Brown-Peterson spectrum $BP$:
\begin{equation}
    MSO_{(3)}\simeq BP\vee\Sigma^8BP\vee\cdots,
\end{equation}
where $H^*(BP;\bZ_3)\cong\mathcal{A}_3/(\beta)$. Since $BP$ is connective, the summand $\Sigma^8BP$ contributes only in degrees eight and above. Thus, in the range $n\leq7$, only the unsuspended $BP$ summand contributes.
We will discuss $BP$ in more detail in Appendix~\ref{BPmethod}.
Therefore, we replace $MSO$ by $BP$, and the $E_2$ page becomes $\mathrm{Ext}_{\mathcal{A}_3}^{s,t}(\widetilde{H}^*(BP\wedge B\bZ_3;\bZ_3),\bZ_3)$.

Now, we consider the Milnor basis of the Steenrod algebra.
Since the Steenrod algebra is a Hopf algebra, we can consider its dual, called the dual Steenrod algebra.
Its algebra structure is explicitly given as follows:
\begin{equation}
    (\cA_3)_*\cong\bZ_3[\xi_1,\xi_2,\cdots]\otimes\bigwedge(\tau_0,\tau_1,\cdots),
\end{equation}
where $\wedge$ denotes the exterior algebra,
with  the degree of generators given by
\begin{equation}
    \mathrm{deg}(\xi_i)=2(3^i-1),\quad\mathrm{deg}(\tau_i)=2\cdot3^i-1.
\end{equation}
Then, the cohomology ring of the spectrum $BP$ is rewritten as $H^*(BP;\bZ_3)\cong\cA_3\otimes_E\bZ_3$, where $E=\wedge(Q_0,Q_1,\cdots)$ and $Q_i$ is the dual of $\tau_i$.
Note that this fact corresponds to the vanishing of $\tau_i$'s on the homology side $H_*(BP;\bZ_3)\cong\bZ_3[\xi_1,\xi_2,\cdots]$.
We then obtain the $E_2$ page as following
\begin{equation}
\begin{split}
    &\mathrm{Ext}_{\cA_3}^{s,t}(\tilde{H}^*(MSO\wedge B\bZ_3;\bZ_3),\bZ_3)\\
    &=\mathrm{Ext}_{\cA_3}^{s,t}((\cA_3\otimes_E\bZ_3)\otimes_{\bZ_3}\tilde{H}^*(B\bZ_3;\bZ_3),\bZ_3)\\
    &=\mathrm{Ext}_{E}^{s,t}(\tilde{H}^*(B\bZ_3;\bZ_3),\bZ_3)
\end{split}
\end{equation}
where we first used $(\cA_3\otimes_E\bZ_3)\otimes_{\bZ_3}V\simeq \cA_3\otimes_E V$
and then we change the ring from $\cA_3$ to $E$.\footnote{%
This second step is possible because $\cA_3$ is free as an $E$-module. Therefore the tensor product functor $\cA_3\otimes_E-$ preserve projective resolutions.}
Now what we should do is to construct a resolution of $\tilde{H}^*(B\bZ_3;\bZ_3)$ in terms of $E$, which  is an exterior algebra and makes computations easy.
The degrees of $Q_i$s are $|Q_i|=2\cdot3^i-1$, so in the range of our interest we should use $Q_0$ and $Q_1$.
Note that $Q_0=\beta$ and $Q_1=\cP^1\beta-\beta\cP^1$  which satisfy $Q_0^2=Q_1^2=0$ and $Q_0Q_1+Q_1Q_0=0$,
as can be easily checked by using Adem relations.
Because we know that $H^*(B\bZ_3;\bZ_3)\cong\bZ_3[\beta x]\otimes\wedge(x)$ and 
that $Q_0,Q_1$ actions on $x$ are given by $Q_0x=\beta x$ and $Q_1 x =(\beta x)^3$, 
we can obtain the free resolution as an $E$-module of $\tilde H^*(B\bZ_3;\bZ_3)$ as follows:
\begin{equation}
\begin{split}
    0\gets\tilde{H}^*(B\bZ_3;\bZ_3)&\gets\Sigma E\oplus\Sigma^3E\oplus\Sigma^5E\oplus\Sigma^7E\oplus\Sigma^9E\oplus\cdots\\
    &\gets\Sigma^6E\oplus\Sigma^8E\oplus\Sigma^{10}E\oplus\Sigma^{12}E\oplus\Sigma^{14}E\oplus\cdots\\
    &\gets\Sigma^{11}E\oplus\Sigma^{13}E\oplus\Sigma^{15}E\oplus\Sigma^{17}E\oplus\Sigma^{19}E\oplus\cdots.
\end{split}
\end{equation}
Because we concentrate on the degree $\le7$, the Adams chart becomes
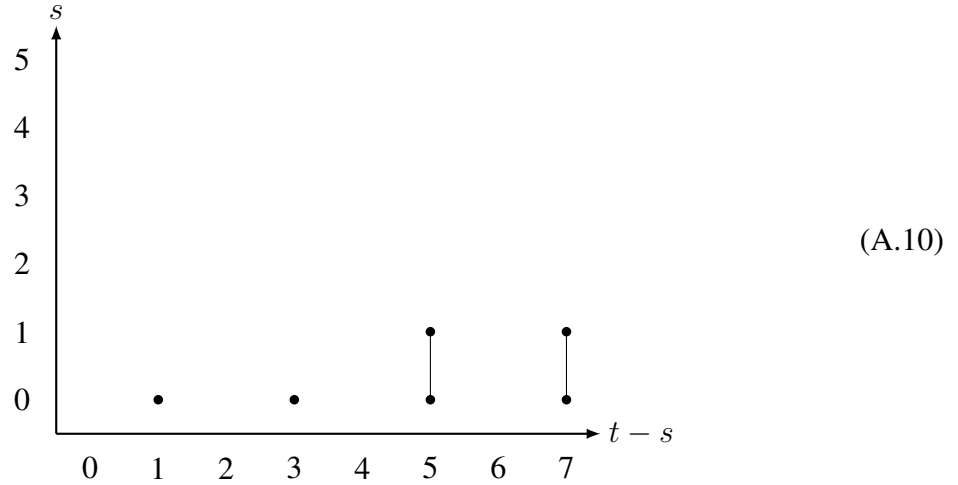
\begin{equation}
\vcenter{\hbox{
\begin{tikzpicture}[scale=0.9]
    \node at (0,-1) {0};
    \node at (1,-1) {1};
    \node at (2,-1) {2};
    \node at (3,-1) {3};
    \node at (4,-1) {4};
    \node at (5,-1) {5};
    \node at (6,-1) {6};
    \node at (7,-1) {7};
    \node at (8.1,-0.5) {$t-s$};
    \node at (-1,0) {0};
    \node at (-1,1) {1};
    \node at (-1,2) {2};
    \node at (-1,3) {3};
    \node at (-1,4) {4};
    \node at (-1,5) {5};
    \node at (-0.5,5.7) {$s$};
    \draw[-latex,thick] (-0.5,-0.5) -- (-0.5,5.5);
    \draw[-latex,thick] (-0.5,-0.5) -- (7.5,-0.5);

    \def\x{0}
    \def\s{0.7mm} 
    
    \fill[black] (1-\x,0) circle (\s);
    \fill[black] (3-\x,0) circle (\s);
    \fill[black] (5-\x,0) circle (\s);
    \fill[black] (7-\x,0) circle (\s);
    \fill[black] (5-\x,1) circle (\s);
    \fill[black] (7-\x,1) circle (\s);

    \draw[-] (5,0) -- (5,1);
    \draw[-] (7,0) -- (7,1);
    
\end{tikzpicture}
}}
\end{equation}
Due to the sparseness of the chart, there is no more differentials, and $E_2\simeq E_\infty$.
The vertical line is the action of $Q_0\in\mathrm{Ext}_E^{1,1}(\bZ_3,\bZ_3)$,
which is often also called $h_0$ and represents the multiplication by $3$.
There is a non-trivial extension in the $t-s=5$ column, so we conclude that the bordism $\tilde{\Omega}^{SO}_5(B\bZ_3)$ is isomorphic to $\bZ_9$.
We also found that $\tilde{\Omega}^{SO}_7(B\bZ_3)$ is similarly given by $\bZ_9$.

\subsection{Using the spectrum $BP$}
\label{BPmethod}
Here we use the Brown-Peterson spectrum $BP$ to solve the extension problem.
The  spectrum $BP$ was first introduced in \cite{BROWN1966149}. For us, its importance lies in the fact that the localization of the $MSO$ spectrum at an odd prime decomposes as a wedge sum of shifts of $BP$,
as proved in the same reference in \cite{BROWN1966149}, using a result of Milnor \cite{MilnorSteenrod}.
At $p=3$, $MSO$ decomposes as  $MSO_{(3)}\simeq BP\vee\Sigma^8BP\vee\cdots$.
Therefore, in the range of interest to us, $MSO_{(3)}$ may be regarded as homotopy equivalent to a single $BP$.
Thus, from now on, we consider only $BP_*$ and $BP_*(B\bZ_3)$ instead of the oriented bordism.

We begin by briefly reviewing the facts about the $BP$ spectrum that will be used in what follows\cite{JW85,Hanke_2016}.
The ring structure of $BP_*$ is given by
\begin{equation}
    BP_*\cong\bZ_{(p)}[v_1,v_2,\cdots],
\end{equation}
where the homological degree of the generator $v_i$ is $2p^i-2$.
Similarly, \begin{equation}
    BP^*\cong\bZ_{(p)}[v_1,v_2,\cdots],
\end{equation}
where the cohomological degree of the generator $v_i$ is $-(2p^i-2)$.

To explain the significance of these generators, let us for the moment consider
general complex-oriented spectra $E$. Here, a spectrum is called $S$-oriented
for a tangential structure $S$,
if it receives a morphism $MS\to E$ from the Thom spectrum $MS$ for the tangential structure $S$.
Then there is a Thom class in $E$-cohomology for any vector bundle with $S$ structure,
and the Thom isomorphism theorem works.
A complex orientation is then an orientation for the stable complex structure.
For more on orientations and Thom isomorphisms in generalized (co)homology theory, see e.g.~\cite{Rudyak},
and on complex orientations and $BP$ in particular, see e.g.~\cite{Ravenel}.

For $E$  complex-oriented, the $E$-cohomology of $\bC P^N$ is 
\begin{equation}
    E^*(\bC P^N)\cong E^*[x]/(x^{N+1}).
\end{equation} This can be proved inductively using the fact that $\bC P^N$ is the Thom space
of the tautological bundle $L\to \bC P^{N-1}$.
Taking the limit $N\to \infty$, we have \begin{equation}
    E^*(\bC P^\infty)\cong E^*[[x]].
\end{equation}
Here, $\bC P^\infty\simeq BU(1)$ is the classifying space for $U(1)$,
and $x=c_1(L)\in E^2(BU(1))$ is the 1st Chern class 
of the universal tautological bundle $L\to BU(1)$
in $E$ cohomology.
Dually, the $E$-homology of $\bC P^N$ can also be determined inductively,
 and  is given by \begin{equation}
    E_*(\bC P^N)\cong E_*\langle 1, b_1, b_2,\cdots b_N\rangle,
\end{equation}  where $b_i\in E_{2i}(\bC P^N)$
are the dual basis to $x^i$, i.e.~they have
the Kronecker pairing   $\langle x^i,b_i\rangle=1$.
This in particular means that $x\cap b_i=b_{i-1}$.
The $E$-homology of $\bC P^\infty$ is obtained by taking the limit $N\to \infty$.\footnote{%
As $MSO$ is itself complex oriented,
this fully determines $\Omega^{SO}_*(BU(1))$.
This in itself might be of some interest.
}

Now, the $p$-th power map $p: U(1)\to U(1)$
leads to the map
\begin{equation}
    BU(1)\xrightarrow{p} BU(1).
\end{equation}
The pullback of $x$ along the map $p$ then has the form
\begin{equation}
[p](x):=p^*(x)= \sum_{i=0}^\infty a_i x^i, \qquad a_i\in E^{2-2i}.
\end{equation} For $E=H\bZ$, this reduces to the standard formula that $c_1(L^{\otimes p})=p c_1(L)$,
and it is known that $a_0=p$ for general $E$.

These data can be used to determine $E_*(B\bZ_p)$ in general,
if we further assume that the coefficient ring $E_*({\rm pt})$ is concentrated in even degrees.
Recall that $B\bZ_p$ is the sphere bundle of $L^{\otimes p}\to \bC P^\infty$.
The homology Gysin sequence then has  the form \begin{multline}
0=E_{2i}(B\bZ_p) \to 
E_{2i}(\bC P^\infty) \xrightarrow{[p](x)\cap } E_{2i-2}(\bC P^\infty) \\
\to E_{2i-1}(B\bZ_p)  \to E_{2i-1}(\bC P^\infty)=0,
\end{multline} where we assumed $i>0$,
the second homomorphism is the cap product with $[p](x)$,
and the vanishing of the first and the last term follows easily from the AHSS and the assumption
that $E_*$ is concentrated in even degrees.

This means that $E_{2i-1}(B\bZ_p)$ is the cokernel of $[p](x)\cap$.
Denoting the image of $b_{i-1}\in E_{2i-2}(\bC P^\infty)$ in $E_{2i-1}(B\bZ_p)$ by $z_i$,
we see that $E_*(\bC P^\infty)$ is generated by $z_i$ over $E_*$, with relations
given by $[p](x) \cap z_i=0$ where $x\cap z_i = z_{i-1}$. More explicitly, we have
\begin{equation}
\label{eq:BP_relation}
    \sum_{i=0}^{m-1}a_iz_{m-i}=0.
\end{equation}

For example, when $E=H\bZ$, this reproduces the standard fact that $H_{2i-1}(B\bZ_p;\bZ)=\bZ_p$
and $H_{2i}(B\bZ_p;\bZ)=0$.
Our main interest is when $E=BP$. For this we need the formula for $[p](x)$ in this case,
which is known to have the form
\begin{equation}
\label{eq:pullback_x}
    [p](x):=p^*(x)=\sum_{i=0}^\infty v_ix^{p^i}=px+v_1x^p+v_2x^{p^2}+\cdots
\end{equation}
which explains the significance of the generators $v_i$ of $BP_*$.

As already mentioned, $MSO_{(3)}\simeq BP$ for $p=3$ in the low range we are interested,
and the first few relations \eqref{eq:BP_relation} are
\begin{subequations}
\label{eq:BP_4relation}
\begin{align}
    3z_1&=0, \\
    3z_2&=0, \\
    3z_3+v_1z_1&=0, \label{eq:BP_Omega5}\\
    3z_4+v_1z_2&=0. \label{eq:BP_Omega7}
\end{align}
\end{subequations}
The first two equations correspond to the facts $\Omega^{SO}_1(B\bZ_3)\cong\bZ_3$ and $\Omega^{SO}_3(B\bZ_3)\cong\bZ_3$, respectively.
The last two solves the extension problems in degrees five and seven in the homological AHSS \eqref{eq:AHSS_sec2.4}.
For example,  the element $z_3\in BP_5(B\bZ_3)\simeq \Omega^{SO}_5(B\bZ_3)$ 
has order $9$, since $3z_3=-v_1z_1\neq 0$ while $9z_3=-3v_13z_1=0$.
This means that the short exact sequence \begin{equation}
 0\to E^\infty_{1,4}{}_{\simeq\bZ_3}\to\Omega^{SO}_5(B\bZ_3)\to E^\infty_{5,0}{}_{\simeq\bZ_3}\to0 
\end{equation}
is non-split.

We can explicitly describe the manifolds and bundles corresponding to the generators,
using $MSO_{(3)} \simeq BP$ in the low degrees.
These follow in a mostly straightforward manner.
We list them below up to degree five.
\begin{itemize}
    \item $v_1=[\bC P^2]\in\Omega^{SO}_4$
    \item $z_1=[S^1\to B\bZ_3]\in\Omega^{SO}_1(B\bZ_3)$ with 
$S^1\to B\bZ_3$ induces the quotient $\pi_1(S^1)\cong\bZ\to\bZ_3$
    \item $v_1z_1=[\bC P^2]\cdot[S^1\to B\bZ_3]=[\bC P^2\times S^1\to B\bZ_3]\in\Omega^{SO}_5(B\bZ_3)$
    \item $z_3$ is $[S^5/\bZ_3\hookrightarrow B\bZ_3]\in\Omega^{SO}_5(B\bZ_3)$  up to a shift by 
    multiples of $v_1z_1$ above. As we have fixed $b_i$ only up to the addition of lower filtration terms,
    this is the most we can say.
    Note that this ambiguity does not affect the relation \eqref{eq:BP_Omega5}.
\end{itemize}

\section{The computation of $\tilde{\Omega}^{SO}_8(K(\bZ,4))$}
\label{app:2}

In this appendix, we determine the group $\tilde{\Omega}^{SO}_8(K(\bZ,4))$. 
This can be done if we use the knowledge of the Anderson dual groups we already determined.
In contrast, we need to solve an extension problem if we want to do this
directly using the AHSS for $\tilde{\Omega}^{SO}_8(K(\bZ,4))$.
We would like to explain how it can be done.

\subsection{Using the Anderson dual}
Our first method is to use our knowledge of the Anderson dual group 
$(\widetilde{I_\bZ\Omega}{}^{SO})^*(K(\bZ,4))$ computed in Sec.~\ref{sec:exBG2BF4}.
There, we already showed that \begin{equation}
(\widetilde{I_\bZ\Omega}{}^{SO})^8(K(\bZ,4))=\bZ^2,
\end{equation}
using the non-vanishing of the differential \eqref{d544}.
The same non-vanishing differential implies \begin{equation}
(\widetilde{I_\bZ\Omega}{}^{SO})^9(K(\bZ,4))=0.
\end{equation}
Now, using the universal coefficient theorem for the Anderson dual,
we see that \begin{equation}
\tilde{\Omega}^{SO}_8(K(\bZ,4))=\bZ^2.
\end{equation}

\subsection{Using the compatibility with suspension}
Now let us determine the same group directly using the AHSS for $\tilde{\Omega}^{SO}_8(K(\bZ,4))$.
As we will see, an extension problem arises in the process
which need to be resolved.
Our strategy is to use the result of  $\tilde{\Omega}^{SO}_7(K(\bZ,3))$ determined in \cite{Jia:2026jhj},
and then use the compatibility 
with suspension, following the approach of \cite{Joyce:2025jyd}.

The $E^2$-page $E^2_{p,q} = \tilde  H_p(K(\mathbb{Z},4),\Omega^{SO}_q({\rm pt}))$ is 
\begin{equation}
\vcenter{\hbox{
    \begin{sseq}[entrysize=9mm]{9}{6}
        \ssdrop{\bZ}
        \ssmove 0 4 \ssdrop{\bZ}
        \ssmove 0 1 \ssdrop{\bZ_2}
        \ssmoveto 4 0 \ssdrop{\bZ}
        \ssmove 0 4 \ssdrop{\bZ}
        \ssmove 0 1 \ssdrop{*}
        \ssmoveto 6 0 \ssdrop{\bZ_2}
        \ssmove 0 4 \ssdrop{*}
        \ssmove 0 1 \ssdrop{*}
        \ssmoveto 7 0 \ssdrop{}
        \ssmove 0 5 \ssdrop{*}
        \ssmoveto 8 0 \ssdrop{F}
        \ssmove 0 4 \ssdrop{*}
        \ssmove 0 1 \ssdrop{*}
    \end{sseq}
}}
\label{tab:E2KZ4}
\end{equation}

where $F = \bZ \oplus \bZ_3$. It follows $\tilde \Omega^{SO}_7(K(\bZ,4))= 0$. As for $p+q=8$, $E^\infty_{8,0} =E^2_{8,0} = F = \bZ \oplus \bZ_3$ and $E^\infty_{4,4}=E^2_{4,4}=\bZ$.
Therefore we have 
\begin{equation} \label{eq:KZ4extension}
    0 \longrightarrow  \left(\tilde H_4(K(\bZ,4), \bZ)\right)_{\cong \bZ} \longrightarrow \tilde \Omega^{SO}_8(K(\bZ,4)) \longrightarrow \left( \tilde H_8(K(\bZ,4), \bZ)\right)_{\cong \bZ \oplus \bZ_3} \longrightarrow 0 \,.
\end{equation}

We now compare this with the AHSS for the bordism groups for $K(\bZ,3)$  in Sec.~\ref{sec:KZ3}.
Consider the suspension isomorphisms $\sigma : \tilde H^\bullet(K(\bZ,3),\bZ) \rightarrow  \tilde H^{\bullet+1}(\Sigma K(\bZ,3),\bZ)$, under which the generator $\iota_3\in \tilde H^3(K(\bZ,3),\bZ)$ is mapped to the generator $\sigma(\iota_3)$ of $\tilde H^4(\Sigma K(\bZ,3),\bZ)$.
By properties of Eilenberg-MacLane spaces, we have 
\begin{equation}
    \begin{aligned}
         \tilde H^{4}(\Sigma K(\bZ,3),\bZ) &\xrightarrow{\;\cong\;} [\Sigma K(\bZ,3), K(\bZ,4)] \\
         \sigma(\iota_3) &\longmapsto \chi \,,
    \end{aligned}
\end{equation}
which determines a map $\chi : \Sigma K(\bZ,3)\rightarrow K(\bZ,4) $ up to homotopy and such that it induces
\begin{equation}
    \begin{aligned}
         \chi^*:\tilde H^{4}(K(\bZ,4),\bZ) &\xrightarrow{\;\cong\;} \tilde H^{4}(\Sigma K(\bZ,3),\bZ) \\
          \iota_4 &\longmapsto \sigma(\iota_3) \,.
    \end{aligned}
\end{equation} 
We also have dually induced maps $\chi_*$ on the homology side. Since the reduced bordism is also a generalized homology theory, we have a suspension isomorphism $\tilde \Omega^{SO}_{i}(K(\bZ,3))\xrightarrow{\;\cong\;}\tilde \Omega^{SO}_{i+1}(\Sigma K(\bZ,3))$. It can be composed with $\chi_*$ to get a morphism (which we will still refer to as $\chi_*$ for brevity) between bordism groups 
\begin{equation}
    \tilde \Omega^{SO}_{i}( K(\bZ,3)) \xrightarrow{ \;\cong\;} \tilde \Omega^{SO}_{i+1}(\Sigma K(\bZ,3)) \xrightarrow{ \;\chi_*\;} \tilde \Omega^{SO}_{i+1}( K(\bZ,4)) \,,
\end{equation}
and between spectral sequences 
\begin{equation}
    \chi_*:  E^r_{p,q}(K(\bZ,3)) \longrightarrow  E^r_{p+1,q}(K(\bZ,4))\,.
\end{equation}
The morphism between AHSSs induced by $\chi$
preserves the filtrations and therefore induces a commutative diagram between the corresponding short exact sequences \cite{Joyce:2025jyd}. 
For $K(\bZ,3)$, we saw in Sec.~\ref{sec:KZ3} that the sequence
\begin{equation}  \label{eq:KZ3extension}
    0 \longrightarrow  \left(\tilde H_3(K(\bZ,3), \bZ)\right)_{\cong \bZ} \longrightarrow \tilde \Omega^{SO}_7(K(\bZ,3)) \longrightarrow \left( \tilde H_7(K(\bZ,3), \bZ)\right)_{\cong  \bZ_3} \longrightarrow 0 
\end{equation}
is non-split, and $\tilde H_7(K(\bZ,3), \bZ ) \cong \tilde H_7(K(\bZ,3), \bZ_3)$, dual to  $\tilde H^7(K(\bZ,3), \bZ_3)$ that is generated by $\mathcal{P}^1 \rho_3(\iota_3)$. 

Between the two sequences \eqref{eq:KZ4extension} and \eqref{eq:KZ3extension} there is the commutative diagram\footnote{%
A priori, we should use $E^\infty_{p,q}$ instead of $E^2_{p,q}$ in this diagram,
but it is straightforward to see that $E^\infty_{p,q}\simeq E^2_{p,q}$ in our case.
} 
\begin{equation} \label{eq:commdiag}
    \begin{array}{c@{\,}c@{\,}c@{\,}c@{\,}c@{\,}c@{\,}c@{\,}c@{\,}c}
0&\to&\tilde H_3(K(\bZ,3),\bZ)_{\cong\bZ}&\to&\tilde\Omega^{SO}_7(K(\bZ,3))&\to&\tilde H_7(K(\bZ,3),\bZ)_{\cong \bZ_3}&\to&0\\
&&\downarrow \chi_* &&\downarrow \chi_*&&\downarrow \chi_*&&\\
0&\to&\tilde H_4(K(\bZ,4),\bZ)_{\cong \bZ}&\to&\tilde\Omega^{SO}_8(K(\bZ,4))&\to&\tilde H_8(K(\bZ,4),\bZ)_{\cong \bZ\oplus \bZ_3}&\to&0
\end{array}.
\end{equation}
Here, the leftmost $\chi_*$ is an isomorphism.
As for the rightmost $\chi_*$,
the $\bZ_3$ summands are mapped isomorphically.
To see this, note that this $\bZ_3$ summand in $\tilde H_8(K(\bZ,4), \bZ)$ 
is dual to $\mathcal{P}^1 \rho_3(\iota_4)$,
and $\chi^*(\mathcal{P}^1 \rho_3(\iota_4)) = \sigma \left( \mathcal{P}^1 \rho_3(\iota_3) \right) $.
Therefore, this $\bZ_3$ summand is indeed the image by $\chi_*$ of $\tilde H_7(K(\bZ,3), \bZ)$.

The commutativity of diagram~\eqref{eq:commdiag} forces the lower sequence to be non-split. To see this, let
\begin{equation}
    u' \in  \tilde H_3(K(\bZ,3),\bZ)\cong \bZ
\end{equation}
be a generator, and let $x'\in \tilde\Omega^{SO}_7(K(\bZ,3))$ be a lift of the generator of
\begin{equation}
 \tilde H_7(K(\bZ,3),\bZ)\cong \bZ_3.
\end{equation}
Then $3x'\in F'_{3,4}$, so we can write
\begin{equation}
3x'=m\,u'
\end{equation}
for some $m\in \bZ$. Since the upper sequence \eqref{eq:KZ3extension} is non-split, one has $m\not\equiv 0\pmod 3$.

Now $\chi_*$ sends the generator $u'$ to a generator
\begin{equation}
u\in  \tilde H_4(K(\bZ,4),\bZ)\cong \bZ.
\end{equation}
From the commutativity, $x:=\chi_*(x')$ 
is a lift of the $\bZ_3$-summand of
\begin{equation}
 \tilde H_8(K(\bZ,4),\bZ)\cong \bZ\oplus \bZ_3.
\end{equation}
So we have \begin{equation}
3x = 3\chi_*(x') = \chi_*(3x') = \chi_*(m\,u') = m\,u.
\end{equation}
Thus the same congruence class $m \bmod 3$ determines the $\bZ_3$-by-$\bZ$ part of the lower extension. Since $m\not\equiv 0\pmod 3$, the sequence \eqref{eq:KZ4extension} is also non-split. It then follows that
\begin{equation}
\tilde\Omega^{SO}_8(K(\bZ,4))\cong \bZ^2.
\end{equation}

\section{Determination of the differentials $d_5$}
\label{app:differentials}
In the main text, we only quoted that the differentials $d_5$
relevant for us is given by $\beta \cP^1$ and $\cP^1 \beta$
for the Anderson dual and the Pontryagin dual, respectively.
Here we review how to determine these.

\paragraph{Two-stage truncation:}
To put our focus on the differentials $d_5$ and $d^5$, it is useful to make
the two-stage truncation of the generalized (co)homology theories themselves.
For any generalized (co)homology theory $E_n(X)$ and $E^n(X)$,
there is a sequence of spaces $E_n$ satisfying $E_n \simeq \Omega E_{n+1}$
such that \begin{equation}
E^a(X) = [X,E_a] ,\qquad E_a(X)= \lim_{n\to \infty }\pi_{a+n}(X\wedge E_n).
\end{equation}  
Such $E_n$ are called to form an $\Omega$-spectrum $E$.
Now, we can perform the Postnikov truncation to the first two stages
of each of the space $E_n$, meaning that we keep only the first two nontrivial homotopy groups.
Call them $F_n$. Then they also form an $\Omega$-spectrum $F$, and define
corresponding generalized (co)homology theories $F_n(X)$ and $F^n(X)$.
The AHSS for $F_n(X)$ and $F^n(X)$ is obtained by simply keeping the first two rows
of the AHSS for $E_n(X)$ and $E^n(X)$.

The spectrum for $\Omega^{SO}(X)$ and $\Omega_{SO}(X)$
is usually denoted by  $MSO$, with $MSO_n$  the $n$-th space.
(It is usually just taken to be a spectrum rather than an $\Omega$-spectrum in the literature,
but we take $MSO_n$ to form an $\Omega$-spectrum for the brevity of presentation.)
Then the truncated spectrum $F$ has the $n$-th space fitting in a fibration
\begin{equation}
K(\bZ,n+4) \to F_n\to K(\bZ, n).
\end{equation} 
The Postnikov invariant of the fibration \begin{equation}
F_n\to  K(\bZ,n) \xrightarrow{k} K(\bZ,n+5) \label{FKK}
\end{equation} is specified by an element \begin{equation}
k\in H^{n+5}(K(\bZ,n);\bZ),
\end{equation} which also specifies a stable cohomology operation raising the degree by $5$
and keeping the coefficient $\bZ$: \begin{equation}
k: H^{n}(X;\bZ)\to H^{n+5}(X;\bZ). \label{coho-op}
\end{equation}
This $k$ is closely related to the differentials $d_5$ and $d^5$ we are after.

It is easy to see that $k$ is nonzero.
If $k$ were zero, $F_n$ is simply a product, and the corresponding
homology theory would  simply be $F_n(X)=H_n(X;\bZ)\oplus H_{n-4}(X;\bZ)$.
This in particular means that the differential $d_5$ would be zero, which would contradict
the fact that the map $F_n(X)\to H_n(X;\bZ)$ is not a surjection,
a fact we confirmed in Section~\ref{sec:curious}.

\paragraph{Determination of the differential:}

It requires some more discussions to 
actually determine $k$, and then translate that information 
to the differentials $d_5$ and $d^5$.
We start from the fact that the group of  stable cohomology operations
of the form \eqref{coho-op} was determined e.g.~in \cite{BMT13}
to be  $\bZ_2\oplus \bZ_3$. 
As \cite{Troue} showed, the Postnikov invariant of $MSO$ is never 2-torsion.
Therefore $k$ is in the $\bZ_3$ part, whose generator is our  $\beta\cP^1$.
We already discussed $k\neq 0$,
so $k=\pm\beta\cP^1$.
As we have not been careful about fixing choices coming from
the automorphism of $K(\bZ,n)$ sending the generator $\iota$ to $-\iota$,
this is as much as we can say at this point.

Now, given a space $X$, this Postnikov class $k$ leads to the exact sequence \begin{equation}
\pi_{n+a}(X\wedge F_n ) \to 
\pi_{n+a}(X\wedge K(\bZ,n))\to
\pi_{n+a}(X\wedge K(\bZ,n+5)),
\end{equation} which leads to the exact sequence \begin{equation}
F_a(X) \to H_a(X;\bZ)\to H_{a-5}(X;\bZ),
\end{equation} the second homomorphism of which is the homology differential $d^5$
we are after.

To describe this homology differential, let us instead describe its dual, \begin{equation}
d_5\colon H^{a-5}(X;\bQ/\bZ) \to H^a(X;\bQ/\bZ),
\end{equation} under the Kronecker pairing,
which is also the cohomology differential for the Pontryagin dual of the oriented bordism.
As $k$ is (up to a sign) the composition \begin{equation}
H^{a}(X;\bZ)\xrightarrow{\rho} H^{a}(X;\bZ_3) 
\xrightarrow{\cP^1} H^{a+4}(X;\bZ_3)
\xrightarrow{\beta} H^{a+5}(X;\bZ),
\end{equation} $d_5$ is (up to a sign) the composition 
\begin{equation}
H^{a-5}(X;\bQ/\bZ)\xrightarrow{\beta^\vee} H^{a-4}(X;\bZ_3)
\xrightarrow{(\cP^1)^\vee} H^{a}(X;\bZ_3) 
\xrightarrow{\rho^\vee} H^{a}(X;\bQ/\bZ).
\end{equation}
Here, for a stable cohomology operation \begin{equation}
H^{a}(X;A)\xrightarrow{f} H^{a+m}(X;B),
\end{equation}
 we use the notation \begin{equation}
H^{a-m}(X;B^\vee)\xrightarrow{f^\vee} H^{a}(X;A^\vee)
\label{eq:C13}
\end{equation} for the Pontryagin dual cohomology operation; they satisfy
on an arbitrary oriented manifold $M$ the relation \begin{equation}
\langle [M],  (fx)  y \rangle
=\langle [M],  x (f^\vee  y) \rangle
\end{equation} when the degrees are appropriate.
Then $\beta^\vee=\beta$, $\rho^\vee=\iota$, and $(\cP^1)^\vee = -\cP^1$.
(More generally, the operation $a\mapsto a^\vee$ on the stable cohomology operations
$H^a(X;\bZ_p)\to H^{a+m}(X;\bZ_p)$ is an algebra anti-automorphism 
of the mod-$p$ Steenrod algebra
and is known to be given by the antipode.)
This determines $d_5$ (again up to a sign) as the composition
\begin{equation}
H^{a-5}(X;\bQ/\bZ)\xrightarrow{\beta} H^{a-4}(X;\bZ_3)
\xrightarrow{\cP^1} H^{a}(X;\bZ_3) 
\xrightarrow{i} H^{a}(X;\bQ/\bZ).
\end{equation} 
Fixing the sign so that it agrees with the explicit examples in Section~\ref{sec:2},
we find
 \begin{equation}
d_5 =  i \cP^1 \beta
\end{equation} for the Pontryagin dual.

The differential  $d_5'$ for the Anderson dual then needs to fit in the commuting square \begin{equation}
\begin{array}{ccc}
H^{a-5}(X;\bQ/\bZ) &\xrightarrow{\beta} & H^{a-4}(X;\bZ) \\
d_5 \downarrow && \downarrow d'_5 \\
H^a(X;\bQ/\bZ) & \xrightarrow{\beta} &H^{a+1}(X;\bZ)
\end{array}
\label{eq:C16}
\end{equation}
which fixes it to be \begin{equation}
d'_5 = \beta \cP^1 \rho.
\end{equation}
This was what we wanted to show.

\bibliographystyle{ytamsalpha}
\def\arxivfont{\rm}
\baselineskip=.95\baselineskip
\bibliography{ref}

\end{document}